\documentclass[prd,aps,twocolumn,a4paper,floatfix]{revtex4-2}

\normalsize

\usepackage{graphicx,psfrag,mathrsfs}
\usepackage{slashed}
\usepackage{mathrsfs}
\usepackage{amsmath,amsfonts,amssymb,amsthm}
\usepackage{diffcoeff}
\usepackage{booktabs}
\usepackage{algpseudocode}
\usepackage{algorithm}
\usepackage{hyperref}
\usepackage{url}
\usepackage{comment,cancel}
\usepackage{accents}
\usepackage{ulem}
\usepackage{xcolor}
\usepackage{tensor}
\usepackage[section]{placeins}
\usepackage[shortlabels, inline]{enumitem}
\usepackage{orcidlink}

\diffdef {sr}
{
  op-symbol = \partial
}

\makeatletter
\newcommand{\sbullet}{%
  \hbox{\fontfamily{lmr}\fontsize{.4\dimexpr(\f@size pt)}{0}
    \selectfont\textbullet}}

\makeatother

\begin{document}

\title{
  Adaptive hp-Refinement for Spectral Elements in Numerical Relativity
}

\author{Sarah Renkhoff$^1$\orcidlink{0000-0002-1233-2593}}
\author{Daniela Cors$^1$\orcidlink{0000-0002-0520-2600}}
\author{David Hilditch$^2$\orcidlink{0000-0001-9960-5293}}
\author{Bernd Br\"ugmann$^1$\orcidlink{0000-0003-4623-0525}}

\affiliation{${}^{1}$Theoretical Physics Institute, University of Jena, 07743
  Jena, Germany,\\
  $^2$CENTRA, Departamento de F\'isica, Instituto Superior
  T\'ecnico IST, Universidade de Lisboa UL, Avenida Rovisco Pais 1,
  1049 Lisboa, Portugal}

\begin{abstract}
  When a numerical simulation has to handle a physics problem with a
  wide range of time-dependent length scales, dynamically adaptive discretizations
  can be the method of choice.
  We present a major upgrade to the numerical relativity code \texttt{bamps}
  in the form of fully adaptive,
  physics-agnostic hp-refinement.
  We describe the foundations of mesh refinement in the context of spectral element methods,
  the precise algorithm used to perform refinement in \texttt{bamps},
  as well as several indicator functions used to drive it.
  Finally, we test the performance, scaling, and the accuracy of the code
  in treating several 1d and 2d example problems,
  showing clear improvements over static mesh configurations.
  In particular,
  we consider a simple non-linear wave equation,
  the evolution of a real scalar field minimally coupled to gravity,
  as well as nonlinear gravitational waves.

\end{abstract}

\maketitle

\section{Introduction}

At the heart of many numerical methods for differential equations lies
discretization,
the transfer of a problem posed on a continuum to a finite set of values.
For spatial discretization especially,
many approaches exist,
from equidistant Cartesian sampling for use with finite-difference methods,
to frequency space decompositions on irregularly shaped sub-domains used with finite-element methods.

The choice of discretization has direct impacts on the results of any given simulation,
since the discretization determines the solution space itself.
A poor choice of sample points might fail to resolve high frequency components of the solution,
or it might cause unphysical numerical noise to accumulate.
This poses an inherent challenge to physics applications seeking to resolve a priori unkown solutions,
which may contain features at different scales.

Using a variable resolution offers a solution to this problem.
High resolution can be used in areas where it is required to resolve the physics to a given accuracy,
in particular to resolve features on short length scales, high frequency modes, steep gradients, or small features requiring high precision,
while a lower resolution is used elsewhere,
saving resources. 
If the areas of interest are known beforehand,
for example the center of the domain,
fixed mesh refinement (FMR) can be applied `by hand'.
In a more general setting, however,
automatically detecting features of interest is desirable.
Allowing such a heuristic detection to determine the discretization used is then referred to as adaptive mesh refinement (AMR).
Although originating in finite-element methods,
which are naturally suited to combining elements of different shapes and sizes,
AMR methods are used in other types of methods as well.
  
Developments in numerical relativity were strongly influenced by \cite{BerOli84a},
which describes a framework for using a flexible AMR scheme for finite-difference methods using rectangular boxes,
recursively overlapping each other.
Of particular note is the PAMR/AMRD toolset
used by Choptuik to pioneer the use of AMR in numerical relativity~\cite{PAMR_RNPL,Cho93}
for the study of scalar field collapse in spherical symmetry,
which motivated the first application of AMR in 3+1 dimensions for black holes~\cite{Bru96,Bru97}.
Box-based AMR approaches are used in GRChombo~\cite{CloFigFin15,AndSalAur22} and HAD~\cite{Lie02},
and nested-box AMR is the basis of many numerical relativity codes to this day,
such as \texttt{BAM}~\cite{BruGonHan06}, AMSS-NCKU~\cite{CaoYoYu08}, Cactus~\cite{cactus_web}, and the Einstein Toolkit~\cite{LofFabBen11}.
See also \cite{RadSpeDre21} for a discussion of the challenges involved in Berger-Oliger type AMR.

As an alternative to overlapping grids,
one can subdivide a given domain into non-overlapping grids,
leading to various types of finite element methods,
see for example~\cite{KreSch74} for time-dependent partial differential equations (PDEs).
Examples of this type of refinement for numerical relativity include SpECTRE~\cite{KidFieFou16}, dendro-GR~\cite{FerNeiHir19}, GR-Athena++~\cite{DasZapCoo21}, and Nmesh~\cite{TicLiAdh22}.

Spectral element (SE) methods choose a set of basis functions for each element,
so that the approximate solution is given by an expansion in a finite set of basis functions.
Examples for SE methods include pseudospectral methods and Galerkin methods.
When applying AMR to spectral elements,
an additional distinction must be made.
Refinement in that case is possible both in terms of element size (h-refinement),
or in terms of the order of the spectral series (p-refinement).
Both techniques can be,
and frequently are,
used at the same time.
Different approaches to dynamical meshes include specially constructed meshes coinciding with physical features, as used for example in SpEC~\cite{SziLinSch09}.

Focusing on applications of AMR in numerical relativity, 
experience has shown that the use of AMR of some kind is crucial
in order to tackle various problems of current interest,
such as black hole or neutron star binaries including mergers,
and the study of critical collapse.
Specifically in the latter case,
it is known that solutions near criticality contain features of ever decreasing scale,
which would be inaccessible without the use of progressive mesh refinement
(or in some special cases, adapted coordinate systems).

In this paper,
we present a major technical upgrade to the numerical relativity code \texttt{bamps},
adding hp-refinement to its pseudospectral method.
While simulations of critical collapse are the science driver for these developments,
the theoretical considerations and technical insights are applicable to a much wider class of problems involving time-dependent PDEs.
We describe the refinement algorithm in detail,
as well as the heuristics used to drive it.
The new version of \texttt{bamps} enables us to explore certain critical collapse spacetimes
with unprecedented efficiency and accuracy,
see~\cite{SuaRenCor22} for results on the collapse of gravitational waves.

The paper is organized as follows.
Sec.~\ref{sec:hpref} introduces basic theoretical considerations of convergence and efficiency for hp-refinement.
In Sec.~\ref{sec:bamps} we describe the code used,
including the overall structure in Sec.~\ref{sec:grids},
and the AMR algorithm used in Sec.~\ref{sec-amr}.
The scaling behavior of the code is described in Sec.~\ref{sec:performance}.
Sections~\ref{sec:nonlinear-wave-equation}, \ref{sec:scalarfield}, and \ref{sec:brillwave}
describe the application of the code to solving a nonlinear wave equation,
the collapse of a real scalar field,
and the collapse of gravitational waves,
respectively.

\section{Basics of hp-refinement}\label{sec:hpref}

In this section, we collect general statements about hp-refinement
regarding error estimates, convergence, and efficiency in the context
of high-order spectral element methods. High-order pseudospectral
methods are discussed for example in \cite{Boy01,Kop06,HesGotGot07},
but not with a focus on hp-refinement, while for example
\cite{KarShe05} consider hp-refinement, but not for high-order
elements. Concretely, our focus is on pseudospectral methods with
polynomial order of around 10 or higher, which are applied  
to predominantly smooth solutions. 

\begin{figure}[t!]
\includegraphics[width=1.05\columnwidth]{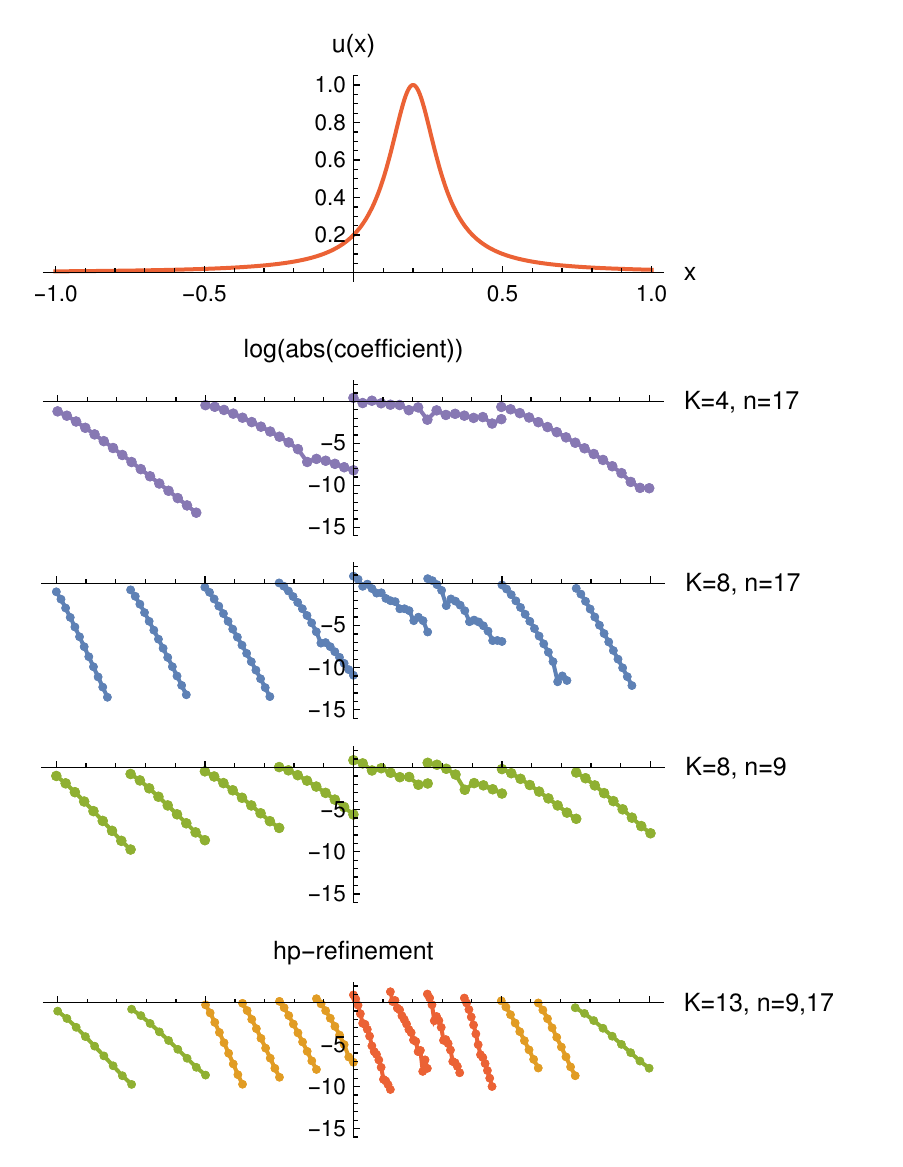}
\caption{
  Example of hp-refinement with spectral elements.
  A function $u(x)$ is discretized on $K$ elements with $n$ collocation points, where $n$ can vary across elements.
  The top panel shows the function $u(x)$ along the $x$-axis,
  while the lower panels show the fall-off of the coefficients of the polynomial expansion in each element on a logarithmic scale.
  The panels are aligned such that the horizontal axes for $x$ and for the fall-off with increasing index
  match for each element.
  In the bottom panel, the result of a specific hp-refinement is shown for a target error of $10^{-7}$.
}\label{fig:hprefexample}
\end{figure}

The goal of AMR and hp-refinement is to optimize efficiency by
adjusting the numerical method locally in space (and possibly also in
time, which we would call hpt-refinement). To prepare for the
discussion of hp-refinement, we show in Fig.~\ref{fig:hprefexample} an
elementary example of the discretization of a function $u(x)$,
here $u(x) = 1/[1+100(x-\frac{1}{5})^2]$, on
spectral elements. There are varying numbers of $K$ elements, and each
element may be discretized by varying numbers of $n$ collocation
points.
For $K=4$, $n=17$, high-order convergence is visible, but the element
near the peak in $u(x)$ shows rather slow but exponential convergence.
For $K=8$, $n=17$, which is one step of h-refinement compared to
the previous panel, errors become smaller but the outer region could
be considered overresolved while the region near the peak is still
somewhat inaccurate.
For $K=8$, $n=9$, there is systematic convergence, but errors are
comparatively large.
In the bottom panel, $K=13$ elements with varying h-levels and
p-levels are used for an efficient representation with a target
absolute error of around $10^{-7}$.

For hp-refinement, we have to define refinement criteria, which
requires defining a measure of the local efficiency, so that we can
optimize the global efficiency. The efficiency of a numerical
discretization can be defined in different ways. For example, (a)
accuracy per number of grid points, (b) accuracy per number of
floating point operations, (c) accuracy per memory usage, or (d)
accuracy per run time. We introduce several of these theoretical
considerations here before turning to a practical implementation in
Sec.~\ref{sec:bamps}.

\subsection{Errors and convergence in a hp-method}

For simplicity, consider scalar functions on the real line.
Consider a numerical approximation $u(x)$ to a sufficiently smooth function
$u_0(x)$ that converges at order $p$ in the grid
spacing $h$, $
  u = u_0 + O(h^p)
$.
More explicitly,
\begin{equation}
u_{h,p}(x) = u_0(x) + c_p(x) h^p + O(h^{p+1}),
\label{eqn:hperror}
\end{equation}
where the coefficient $c_p(x)$ is
a function independent of $h$ (as suggested by a Taylor expansion) and
\begin{equation}
  \epsilon(x;h,p)=c_p(x) h^p
\end{equation}
is the leading order error term.
Here we introduce $h$ as a small parameter that characterizes the
discretization of the domain, the derivatives, and/or the PDE equation,
as opposed to just a displacement in $x$.

For concreteness, we consider a discretization where $h = L/K$, $L$
the length of a 1d interval, and $K$ the number of cells or elements
in the interval, such that each cell has size $h$. For finite difference
(FD) methods, $h$ is the `grid spacing' with one point per cell, and
$p$ is, in particular, the order of the FD approximation of
derivatives. For a spectral element method (pseudospectral (PS), but also
discontinuous Galerkin (DG)), each cell of size $h$ can be discretized by
$n = N+1$ collocation points $x_i$. Polynomials of polynomial
order up to $N$ are used, so that $p = n$ is the typical order of
convergence.  (In this notation, a linear function has $N=1$, and the
term beyond linear is at $n=2$.)

For example, a 2nd-order FD-method with error $O(h^2)$ becomes 4 times
as accurate with 2 times the number of grid points, since $h\sim
\frac{1}{K}$ and error $\epsilon\sim\frac{1}{K^2}$.
For a SE method with error $O(\epsilon)$ for $p$ points per cell, 2
times the number of collocation points implies an error of
$O(\epsilon^2)$, since $\epsilon(p) \sim h^p$, and
$\epsilon(2p) \sim h^{2p} \sim \epsilon(p)^2$.

Consider now an hp-refinement method using spectral elements, where
both $h$ and $p$ can be varied freely. A key question is under which
conditions it is more efficient to decrease $h$ or to increase $p$ in
order to reduce the error.

In terms of the computational degrees of freedom, or simply in
terms of the number of grid points, there are a total of $N_{total}= K
n$ grid points for the interval of size $L$ assuming $K$ equal size cells
with $n$ points each. Given $N_{total} = K n$, we ask the question
how the error changes if we double the degrees of freedom by either
using $2K$ cells with $n$ points or $K$ cells with $2n$ points for the
same total of $2Kn$ points overall. Since $h=L/K$ and $p=n$,
this corresponds to the calculation of $\epsilon(h,p)$ for different
$h$ and $p$. For an example, see Fig.~\ref{fig:hprefexample}

The short answer is, spectral methods always win, or p-refinement
is always more efficient than h-refinement,
{\it assuming certain conditions are met}.
In particular, (i) we consider smooth functions
$u_0\in C^\infty$, (ii) investigate the convergent regime where
$\epsilon\ll1$, and (iii) have infinite numerical precision.

Let us first assume that conditions (i)--(iii) are satisfied and
consider the error estimate given in (\ref{eqn:hperror}).  For the comparison of
different methods, we have to include the coefficients $c_p(x)$ in
the calculation. Restricted to a single location $x$, we have
\begin{eqnarray}
\epsilon(h,p) &=& c_p h^p,
\\
\epsilon(\frac{h}{2},p) &=& c_p (\frac{h}{2})^p = \frac{1}{2^p} \epsilon(h,p),
\\
\epsilon(h,2p) &=& c_{2p} h^{2p} = \frac{c_{2p}}{c_p} h^p \epsilon(h,p).
\end{eqnarray}
This implies for the relative changes
\begin{equation}
\frac{\epsilon(\frac{h}{2},p)}{\epsilon(h,p)} = \frac{1}{2^p},
\qquad
\frac{\epsilon(h,2p)}{\epsilon(h,p)} = \frac{c_{2p}}{c_p} h^p,
\end{equation}
and for the comparison of h- and p-refinement
\begin{equation}
\frac{\epsilon(h,2p)}{\epsilon(\frac{h}{2},p)} = \frac{c_{2p}}{c_p} 2^p h^p.
\end{equation}
p-refinement is more efficient than h-refinement if the last
expression is less than 1. In particular, this is the case if we are
in the `convergent regime' (assuming $\frac{c_{2p}}{c_p}<1$) and if
$2h<1$. Furthermore, for decreasing $h$ the condition on the
coefficients, which are derived from the function $u_0$, becomes less
restrictive. For exponential convergence, we in fact expect
$\frac{c_{2p}}{c_p}\ll1$.
As an aside, doubling $p$ is a rather large step for spectral methods,
but similar relations hold for increasing the order $p$ in increments
of $+1$ or $+2$, etc. We consider doubling here to use the same number
of degrees of freedom in both cases.

With regard to (i), for non-smooth functions $u_0\in C^k$, a spectral
method may lead to algebraic convergence of order (for example) $k+2$
\cite{Boy01}. Other algebraic orders are possible, including
half-integer powers like $k+\frac{3}{2}$. Sufficient smoothness of
$u_0$ is an obvious criterion for the applicability of spectral
methods, while non-smoothness does not automatically rule out spectral
methods, because they still converge. A large class of problems deals
with shocks and conservation laws, which is beyond our discussion here
(see for example \cite{Hes18}).

With regard to (ii), while for large $K$ and $n$ the exponential
convergence is stronger than any polynomial factor, for small $K$ and
$n$ there may be a regime where h-refinement and p-refinement offer
comparable gains. In other words, spectral methods win beyond some
minimal number of grid points. This minimum tends to be rather small,
but depends on the function $u_0$.

Let us assume that the convergent regime for some
functions $u_0$ starts only beyond a specific $n_{conv}$.  For
example, $u_0$ could be constructed from just high order
components in a polynomial basis. Only when $p=n$ is sufficiently
large, $n\ge n_{conv}$, can the error start to decrease
exponentially. A special example would be $u_0(x) = \sin(k x)$, which
is only captured accurately if a Fourier series includes sufficiently
high frequencies. For a single Fourier mode, h-refinement might help,
since on a cell of size $h/2$ fewer cycles have to be resolved, and
relative to the smaller cell short wave lengths have become longer
wave lengths. Another special case is a well-localized wave
packet. Suppose this wave packet is well resolved for $n$ points in a
cell of size $h$. If the same packet is placed in a single cell with
$h'=10h$, then $n'>n$ points are probably needed to resolve the
packet.
The localized wave packet suggests hp-refinements that vary with
position.  A function $u_0$ may be optimally approximated by a pair
$(h,p)$ in some region, where it varies slowly on the scale of
$(h,p)$, while in another region $u_0$ may exhibit high frequency
features that require smaller $h$ and/or larger $p$ for optimal
efficiency.

With regard to (iii), spectral element methods are usually implemented
with finite numerical accuracy. Depending on the calculations required
to, for example, find the approximate solution $u$ of a PDE, round-off errors
may be the dominant, limiting factor for the accuracy of the final
result. This leads to the typical result that a spectral method may
show an exponential drop in the error as $n$ is increased, say down to
$\epsilon \approx 10^{-12}$ for $n\approx 20$, but increasing $n$
further does not decrease the error further, but the error
$\epsilon(n)$ levels off and may even increase for increasing $n$.
From the perspective of hp-refinement criteria, if the round-off floor
has been reached by increasing $n$ for p-refinement, it should be more
efficient to switch to h-refinement. While the overall accuracy may be
the same, computations on two cells of size $h/2$ with $n$ points can
be expected to be more computationally efficient than for a single
cell of size $h$ with $2n$ points. See the discussion of computational
efficiency that follows.

For FD methods, it may be hard to reach this level of round-off error,
while for SE methods reaching round-off may be straightforward, but a
major design objective is to, say, lower the round-off floor from
$10^{-5}$ to $10^{-12}$ by an improvement of the spectral method. A
concrete example are the optimized spectral methods for
certain elliptic problems by Ansorg et
al.\ \cite{MeiAnsKle08}, which achieve this in part by a clever choice
of coordinates.

In conclusion, theoretical estimates for errors and convergence of SE
methods are available. Their applicability, however, depends on the
smoothness of $u_0$, and on application in the convergent regime,
which may require reaching some minimal resolution, as well as
avoiding the numerical round-off for high order schemes.

\subsection{Operation count for hp-refinement}

In a spectral element method for the numerical solution of a PDE, the
most expensive part of the calculation is often the computation of the
numerical derivatives. For a 1d problem with $n$ grid points, the
algebraic (non-derivative) part of the right-hand-side computation
requires typically on the order of $O(n)$ floating point operations,
while computing derivatives can require $O(n^2)$ operations for direct
matrix methods.  In special cases this may be $O(n\log n)$, say for
FFT-Chebychev methods, but since $n$ in many examples is comparatively
small ($n<50$), direct matrix methods for correspondingly small $n$
are more efficient, for example \cite{Boy01,Bru11}. Hence we will restrict
the discussion to the case $O(n^2)$.

In $d$ dimensions, consider a cube with the same number of collocation points in
each direction and
\begin{equation}
V = n^d
\end{equation}
points in total. We define the vector of function values $u_i$ with a
linear index $i=0,\ldots,V-1$. Multiplication of a vector with $V$
elements by a square $V\times V$ matrix is in general a $O(V^2)$
operation. However, let us consider spectral methods for first order
PDEs that involve only the standard partial derivatives $\partial_j$
in each direction. For $d=1$, the $n\times n$ derivative matrix $D$ is
dense (full). For $d=2$, we define sparse derivative matrices
$D_1=I\times D$ and $D_2=D\times I$, which are $n^2\times n^2$
matrices defined by the Kronecker product of $D$ with the $n\times n$
identity matrix. Assuming that the sparsity is utitilized in the
computation, computing derivatives is not an $O(V^2)$ operation, but of order
\begin{equation}
n_{\mathrm{ops}}(n) = O(n V) = O(n^{d+1}).
\end{equation}

Using an estimate for the number of floating point operations in a spectral
method, we can define efficiency as `accuracy per work', or
inefficiency as `work per accuracy', or since accuracy is the
inverse of error (the smaller the error, the higher the accuracy),
inefficiency is error times work.
For the hp-method as described above, with $K$ cells in each dimension,
error times work is
\begin{equation}
\alpha_{\mathrm{ineff}} = O(h^p) n_{\mathrm{ops}}(n) = O(K^{-n}) O(n^{d+1}).
\end{equation}
Comparing SE methods in terms of $\alpha_{\mathrm{ineff}}$ takes the work in
terms of the operation count into account.

For p-refinement, with $K$ and $d$ constant, $\alpha_{\mathrm{ineff}}$ is the
product of an exponential in $-n$ and a polynomial with leading order
$n^{d+1}$. Therefore, assuming that we consider the
regime of exponential convergence for the spectral method, the
exponential reduction of the error outweighs the polynomial increase
of the operation count. Incidentally, since $d$ is constant, this
would also hold if the operation count for derivatives was $O(n^{2d})$
instead of $O(n^{d+1})$.

For h-refinement the work for 1d derivatives per element remains
constant, while for p-refinement the work for 1d derivatives
increases, which may be compensated by faster convergence.
For h-refinement with a factor 2, the operation count is
$2^{d}n_{\mathrm{ops}}(n)$, 
when ignoring overhead at cell interfaces, while the error
decreases by a factor of $1/2^{n}$. The overall gain in efficiency
(reduction in inefficiency) is $1/2^{n-d}$.
For p-refinement by a factor 2, the operation count increases by a factor
of order $2^{d+1}$, while the error decreases by a factor
$\frac{c_{2p}}{c_p}h^p$.

For hp-refinement, h-refinement is cheaper in terms of additional
operations, but overall p-refinement is still favored in the regime
defined by (i)--(iii) in the previous section. In practical
applications, some measure of the work should be included, and the
balance between h- and p-refinement can then be based on the actual values of
$n$, $K$, and the work estimate.

Similar considerations hold for specific time-stepping algorithms. On
the one hand, clustering of points on spectral elements may require
smaller time steps, for example for explicit Runge-Kutta
time-stepping. On the other hand, this is rewarded with smaller errors
in the time discretization, which in turn affects the
work and accuracy balance of hp-refinement.

\subsection{Memory usage of hp-refinement}

Another aspect of efficiency is memory usage. FD and SE methods often
have comparable memory usage of order $O(V)$. In particular, the
additional storage for differentiation matrices is often small compared to
the storage of $O(n_{var} V)$ function values for $n_{var}$
variables. We can ask which method requires the least resources to
achieve a fixed error bound, say maximum pointwise error of
$\epsilon=10^{-9}$. Considering hp-refinement for such an error bound,
in principle we can also balance h-refinement and p-refinement to
minimize $V$. 

However, in many of our applications memory (RAM) size limitations of hardware
are not an issue. With \texttt{bamps}, we rarely perform time evolutions that
use the maximum of memory available per node, rather we utilize more nodes than
required for memory to gain access to more CPUs for faster execution.
Employing more nodes also implies higher memory bandwidth
for accessing the same total memory.

\subsection{Run time of hp-refinement}

Having discussed convergence with $h$ and $p$, operation counts for
typical hp-methods, and memory constraints, we turn to another
important metric for performance: How quickly does the code run? In
particular, for large simulations on supercomputers, the bottom line
may be how many CPU hours are required.

While the number of floating point operations required to complete a
simulation is a relevant metric, different methods implemented by
different codes and run on different hardware typically show run times
that are not trivially correlated with flops (floating point
operations per second). This is not that surprising, because in a
complicated code like \texttt{BAM} or \texttt{bamps} solving complicated problems
(Einstein equations) there are many non-trivial issues like
maintaining optimal load of floating point units (including vector
units like AVX etc.), parallelization, and memory access.

As a key example, even when considering just the right-hand-side
calculation (and not complications of AMR or parallelization) of the
Baumgarte-Shapiro-Shibata-Nakamura (BSSN) or generalized harmonic gauge (GHG) formulations, \texttt{BAM} and \texttt{bamps} seem to be significantly bound
by memory access speed, rather than by the flops achievable by the
hardware. Typical (3+1)-dimensional simulation codes are often
memory-bound rather than compute-bound (in significant parts of the
calculation). Consider {arithmetic intensity}, that is the number of
floating operations performed on each byte read from RAM into the CPU,
or work per memory traffic.
For current CPU/RAM platforms, and for the large
number of 3d variables in a typical BSSN or GHG calculation, the
arithmetic intensity is often comparatively low, so that the memory
channels are saturated, while the CPUs/FPUs are not.  Note that this
is the case even when using differentiation matrices in our spectral
methods. SE tend to be more compute intensive than FD, but in typical
examples even the \texttt{bamps} code is in part memory-bound, and not
compute-bound. See \cite{Bru11} on \texttt{bamps}, where there are indications
that the strong performance gain on a GPU compared to a CPU can be
attributed mostly to the much faster memory interface of the GPU,
rather than the increased flops for the GPU.

In conclusion, for dynamic AMR with hp-refinement it is worthwhile to
include either offline or live benchmarks that measure the speed of
execution for different parts of the code.
While the theoretical          
considerations above can be a good guideline for some aspects of
performance, the optimal balance of h- and p-refinement to reach a
certain error criterion should also consider run-time
benchmarks. Furthermore, benchmarks of this type can be helpful for
load balancing of parallel hp-refinement.

\section{The bamps Code}\label{sec:bamps}

This section first gives a high-level overview,
before describing specific parts of the algorithm in detail,
such as refinement indicators and load balancing.

\subsection{Grid Setup}\label{sec:grids}

In \texttt{bamps}, the numerical domain is organized in a hierarchical
structure.  The full total of the domain, is divided into up to 13 different
`patches', each patch being defined by a patch type (cube, spherical shell,
or transitional patch), and a direction (positive X, negative Y, etc.),
corresponding to its position and orientation on the overall domain. Each
patch is constructed as a cube in patch-local coordinates $(u, v, w)$.
Depending on the patch type, specific coordinate transformations are used to
map the patch-local coordinates to a global set of Cartesian coordinates $
(x, y, z)$.  These transformations are constructed such that they match at
the boundaries between patches.  Finally, boundaries of patches with
different orientations are connected as to form an overall spherical domain,
referred to as a `cubed sphere'
\cite{RonIacPao96}.  The particular construction used in
\texttt{bamps} is described in more detail in \cite{HilWeyBru15}.

Each patch contains any number of `grids', which are self-contained
spectral elements.  On each grid, every spatial dimension is
discretized using a nodal spectral grid of points, either using
Chebyshev-Gauss-Lobatto or Legendre-Gauss-Lobatto collocation points
in the $(u, v, w)$ coordinates.

For the present work, we consider parallelization with MPI (Message
Passing Interface). In normal parallelized operation, a variable list
of grids is distributed across a fixed number of MPI processes. Each
MPI process contains roughly the same number of grids (see
Sec.\ref{sec-load-balancing} on how this is achieved).
Each process stores its grids and their metadata in its own copy of a singleton data structure,
which also contains global properties of the domain.
While some metadata is used to track
grids not local to a particular process, no part of the state vector is
duplicated between different processes.

Since the publication of \cite{HilWeyBru15}, several changes to the
grid structure have been made, most notably symmetry boundaries,
that is those boundaries of the domain on which symmetry conditions are
enforced, are located inbetween grids, rather than being implemented
using `half grids' overlapping the symmetry plane.

This necessitates more care to preserve the parity of the solution,
since without extra steps the filtering mechanism used to prevent the
growth of unphysical high frequency modes will lead to parity
violations, which in some cases seems to lead to unstable behaviour at
the outer boundaries.  We partially counteract this by providing such
boundaries with virtual neighbor grids, containing data that exactly
fulfills the parity conditions.  We can then apply the same penalty
method used between all internal grids as a boundary condition.
Additionally, parity conditions on derivatives are generally enforced
explicitly by setting the derivatives of fields with even parity to
zero at the boundary.

\subsection{Adaptive Mesh Refinement}\label{sec-amr}

The general grid structure described above serves as the base mesh, on
which AMR is then applied.  Using the algorithm described in
Sec.~\ref{sec-amr-algorithm}, we recursively generate new grids in
order to supply additional resolution where needed to adequately
represent the solution, and consolidate these fine grids back into
fewer coarser ones once the additional resolution is no longer
required (h-refinement).  The per-grid resolution is also adjusted
according to a separate refinement indicator (p-refinement).

To determine which areas of the domain require refinement or coarsening,
we employ different indicator functions,
as described in Sec.~\ref{sec-amr-indicators},
which are evaluated periodically during the evolution,
typically every 100 timesteps.

\subsubsection{Data Structures}\label{data-structures}
  
A popular data structure for h-refinement in three dimensions are
oct-trees (or octrees), for example \cite{Kho97,KidFieFou16,FerNeiHir19,DasZapCoo21},
which are natural for recursive, local domain decomposition. A given
grid, the `parent' grid, is subdivided in each spatial direction by a
factor of two, resulting in 8 (in 3d), 4 (in 2d), or 2 (in 1d) `child'
grids. Given a set of root grids, which in \texttt{bamps} corresponds
to an initial grid configuration based on coordinate patches, each
child grid has a unique parent grid, and the data structure also keeps
track of neighborhood relations between grids.

In essence, \texttt{bamps} implements a set of distributed,
parallelized trees of grids, which is similar, for example, to the
forest of octrees in \cite{BurWilGha11}. Incidentally, some versions
of the numerical relativity code \texttt{BAM} starting with
\cite{BruTicJan03} were internally based on octrees as well, which
provided experience with a prototype for a MPI-parallel octree
implementation. However, fully local mesh refinement was rarely used,
rather the octree was configured for the nested, moving box algorithm
for compact binaries as in \cite{BruGonHan06}, and later replaced by a
more efficient box-based algorithm for large, nested boxes
\cite{ThiBerBru11}.

For \texttt{bamps}, we decided to explore a non-standard
implementation of octrees, where the data structures do represent a
virtual tree, but the actual implementation is directly based on lists
and local list operations. Assuming familiarity with elementary data
structures like lists and linked lists, a tree is a specific graph of
nodes with links between parent and child nodes. A binary tree is an
efficient way to store and retrieve data in an ordered list of
nodes. For AMR, the ordering is given by the geometry of the domain
decomposition in 3d (or 2d, 1d). Since for PDEs a key operation is the
exchange of information between neighboring nodes across grid
interfaces,
for convenience and efficiency an implementation may also store links (pointers) between nodes and their neighbors (`siblings' and `cousins'),
see for example the `fully threaded trees' in~\cite{Kho97},
even though some of this information is redundant and can be deduced from the parent/child links.
An alternative to linked lists and linked trees is based on hash-based
node identification, which can be combined efficiently with the
concept of space-filling curves~\cite{GriZum99a,Zum12}.

In \texttt{bamps}, we do not implement a general purpose octree, but
guided by the actual requirements of parallelized, adaptive
hp-refinement we arrived at the following, in some aspects simpler
model of a list of grids.
The construction is based on the following observations and
application specific simplifications.

First, the physical domain is covered by a collection of elements (or
grids), which for the purpose of parallelization is organized as a
global, ordered list corresponding to a space-filling curve, see the
discussion in Sec.~\ref{sec-load-balancing}. 
By ordering the elements in this way,
each can be uniquely identified by its position, or index, in the list.
This makes it possible to encode information about neighborhood relations between elements simply by storing the list indices of neighboring elements.
We choose to maintain this ordered list of grids directly in all AMR operations.

Second, notice that AMR operations like refining and coarsening
correspond to local list operations, assuming a well-formed octree and
z-ordered space-filling curve.  In particular, creation of child grids
corresponds to replacing a single node in the list by several nodes,
which by construction of the z-ordered curve for an octree places the
new elements next to each other in the list. Coarsening means
replacing several child grids by a single (parent) grid, which again
is a local operation in the z-curve, in particular since only children
without children of their own can be removed.
This allows us to maintain the congruency of all stored list indices
by adding and subtracting precomputed offsets to and from stored indices,
based on the refinement and coarsening operations of each element and its neighbors.

Third, given a list of $n$ elements stored contiguously in memory,
inserting and removing $n$ elements is potentially an order $O(n^2)$
operation. For the h-refinement considered here, however, we build the
list of refinement flags ahead of time, and the construction of the
new list can then be performed by a single sweep of $O(n^1)$
operations.
In practice, two sweeps are required,
one to determine new element indices,
and one to assemble to new array,
but overall
this is still a $O(n)$ operation.

While this method of a global list, implementing
`a-tree-without-a-tree', works efficiently, see
Sec.~\ref{sec:performance} on the performance and scaling of
\texttt{bamps}, we leave it to future work to investigate whether
there are significant differences in performance and/or simplicity
compared to other octree implementations.

\subsubsection{Algorithm}\label{sec-amr-algorithm}

The AMR procedure consists of several steps:
\begin{enumerate}
  \item Evaluate the h-refinement indicator function on each grid.
  \item Generate a set of initial h-refinement flags.
  \item Modify the h-refinement flags to satisfy refinement constraints.
  \item Apply h-refinement operations.
  \item Perform load balancing to consolidate grids that are marked to be coarsened together.
  \item Apply h-coarsening operations.
  \item Evaluate the p-refinement indicator function on each grid.
  \item Apply p-refinement and p-coarsening operations.
  \item Apply final load balancing.
\end{enumerate}

We choose to use refinement indicators that are purely grid-local functions,
and as such can be evaluated by each MPI process on all local grids without the need for inter-process communication.
This precludes indicators based on large scale feature detection,
or preemptive refinement based on indicator values of neighboring grids.

The generated indicator function values are then compared to an interval of values deemed acceptable.
This interval is set by the user as an external parameter,
and appropriate values depend strongly on the chosen indicator function.
For example, 
the interval $[10^{-12}, 10^{-9}]$ has been found to give good results
for the truncation error estimator (Sec.~\ref{sec-error-indicator}).
Suitable estimator bounds for the smoothness heuristic (Sec.~\ref{sec-smoothness-indicator}) depend strongly on the equations and quantities being evolved.
See Tab.~\ref{tab:indicator-values} for examples.

\begin{table}
\caption{Examples of refinement indicator settings that are known to generate useful amounts of refinement for evolving linear and nonlinear wave equations, as well as the Generalized Harmonic Gauge (GHG) formulation of GR.}\label{tab:indicator-values}
\begin{tabular}{lll}\toprule
& Smoothness & Truncation Error \\\midrule
Wave eq.\ (linear) & $[0.005, 0.05]$ & $[10^{-12}, 10^{-9}]$ \\
Wave eq.\ (nonlinear) & $[0.001, 0.01]$ & $[10^{-12}, 10^{-10}]$ \\
GHG (Kerr) & $[0.001, 0.01]$ & $[10^{-12}, 10^{-9}]$ \\
GHG (Brill wave) & $[0.001, 0.005]$ & $[10^{-15}, 10^{-9}]$ \\
GHG + scalar field & $[0.001, 0.0025]$ & $[10^{-12}, 10^{-9}]$ \\\bottomrule
\end{tabular}
\end{table}

If the indicator value is above the allowed maximum,
the grid is flagged for refinement.
Vice versa, if the value is below the set minimum,
it is flagged for coarsening.
Typically, the smoothness heuristic is chosen as the h-refinement indicator,
to make sure each grid represents a sufficiently small part of the solution as to be smooth.
If this cannot be achieved,
at least the use of this indicator should contain any non-smooth parts of the solution in as small a region as possible,
and help to preserve the overall quality of the solution.
Our notion of `smoothness' here is not related to the smoothness of the continuum solution, but rather to the quality of its numerical approximation.
The above procedure will assign each grid a `target h-level' that falls within $\pm1$ of its current h-refinement level.
Note that while each application of the AMR algorithm will only raise or lower the refinement level of a given grid by one,
it can be applied iteratively until all indicator bounds are satisfied on all grids.

\begin{figure}[t!]
\includegraphics[width=0.8\columnwidth]{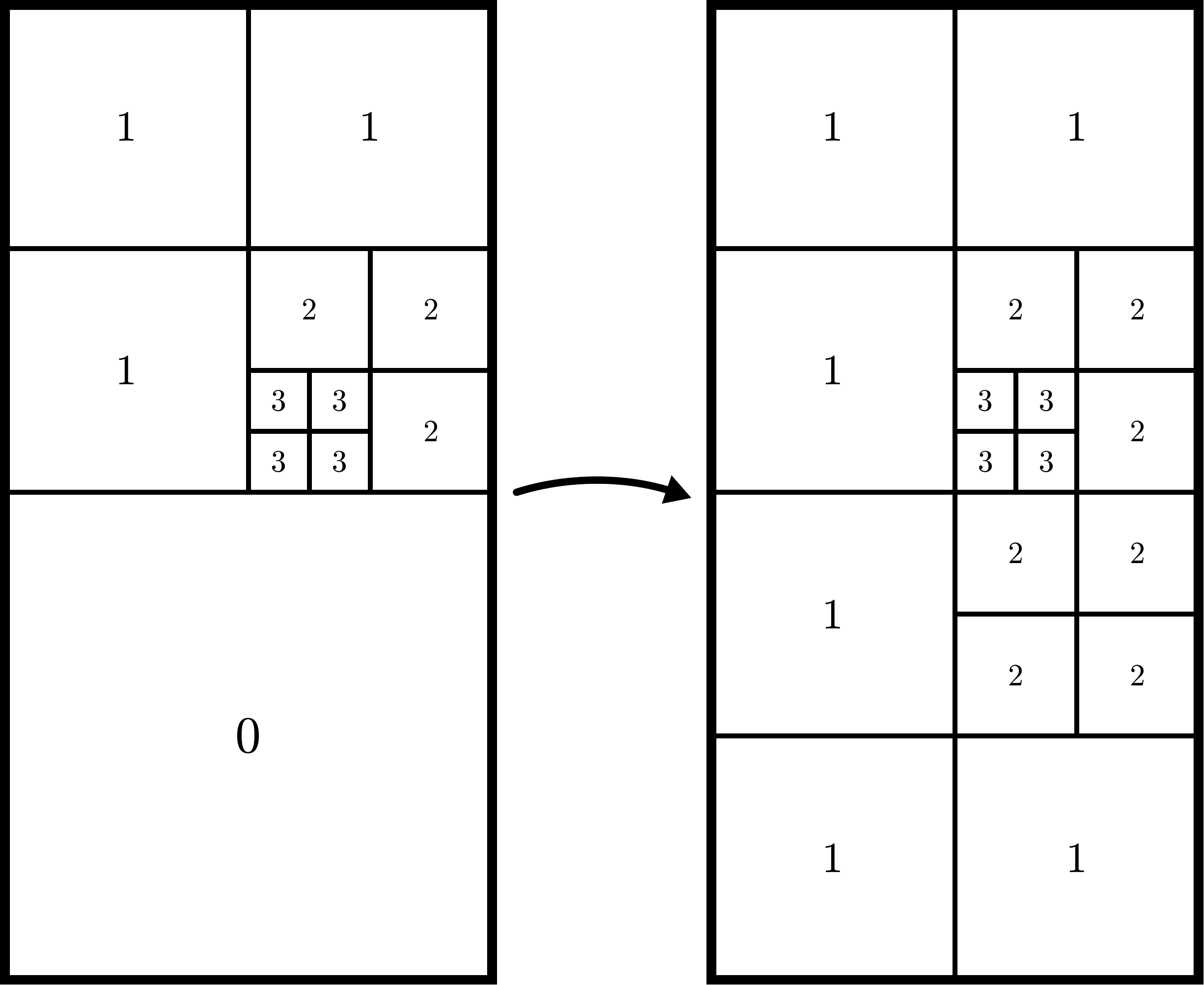}
\caption{A grid structure that conforms to an octree, but is not compatible with the 2:1 refinement condition, and a modified structure which does, demonstrating the resulting refinement propagation, both with labels denoting the refinement level.}\label{fig:illegal-legal-grid}
\end{figure}

As the next step requires information about the flags of neighboring grids,
which might be stored on different MPI processes,
these initial h-refinement flags are then synchronized across all MPI processes.
The flags generated by this procedure are then modified to ensure that the end state satisfies the constraints for a `legal' \texttt{bamps} grid.
As illustrated in Fig.~\ref{fig:illegal-legal-grid},
we impose a 1:2 condition on the grid structure,
meaning that in crossing any boundary between grids,
the h-refinement level may only change by 0 or $\pm1$:

\begin{algorithm}[H]
\caption{Satisfying the 2:1 condition for h-refinement}\label{alg:1-2}
\begin{algorithmic}
  \State $l \gets l_{\mathrm{max}}$
  \While{$l \ge 0$}
    \For{each $grid$}
      \For{each $neighbor$ of $grid$}
        \If{$neighbor.level < l-1$}
          \State $neighbor.level \gets l-1$
        \EndIf
      \EndFor
    \EndFor
    \State $l \gets l-1$
  \EndWhile
\end{algorithmic}
\end{algorithm}

This constraint makes the structure of possible grids conform to the leaves of an octree.
Since grids do not overlap,
only the leaves of that tree actually exist as grids.
However, coarsening operations require knowledge of which groups of grids correspond to `siblings' in the virtual grid tree.
For this reason, each grid is assigned a unique sequence of numbers that mark its position in the tree,
for example a grid marked with the sequence $\{0, 3, 2\}$ is the 2nd child of the 3rd child of a parent grid with the id $0$.

At the same time, refinement is given precedence over coarsening and non-refinement.
In combination, these principles lead to potentially propagating refinement into grid regions that were not originally flagged as needing refinement.
Since coarsening is given the least precedence,
it will only be applied if an entire group of sibling grids,
meaning grids that share a parent node in the virtual grid octree,
is flagged for coarsening.
Otherwise, the coarsening flags are disregarded.

After the refinement flags have been modified to satisfy the constraints,
all h-refinement operations can take place.

To perform h-refinement on a grid, $2^d$ new grids are created,
with $d$ being the dimension of the domain,
to fill the region currently represented by their parent grid.
The solution is then interpolated from the parent grid onto its children,
using Lagrange interpolation with barycentric weights \cite{BerTre04}.
Finally, the metadata about the parent's neighbor grids is transferred to the generated children.
This includes their position in the global grid list, their level, the patch they belong to,
as well as information about size and orientation of neighboring grids.
The new metadata depends on both the current refinement level,
and the refinement flags of any neighboring grids.
A set of nested lookup tables is used to generate the metadata that corresponds to the grid state after all refinement and coarsening has taken place.

Because refining a grid into several children is always a process-local operation,
and it generates new grids that must be taken into account for and potentially moved during load balancing,
all refinement operations can and must happen before load balancing.

Similarly, because coarsening a group of grids requires all grids involved to be local to a single process,
coarsening operations must happen after a pass of load balancing,
during which grids flagged for coarsening are shifted to consolidate all sibling grids on the same MPI process.
To facilitate this,
the grid weighting system of the load balancing procedure is used.
Within a group of grids flagged for coarsening,
all but one are assigned a weight of $0$,
since they will cease to exist.
The other grids are assigned a weight depending on their individual resolution (see Sec.~\ref{sec-load-balancing}).
This results in such sibling grids always being assigned to the same grid list segment.

The coarsening procedure functions like the refinement procedure in reverse;
one new grid is created and data from the old grids is interpolated onto it.
Like in the refinement step,
all metadata from the old grids must be consolidated and modified to reflect the end state of the AMR operation.
Here, too, a set of nested lookup tables is used to generate the correct metadata.

Once all h-refinement operations are complete,
the refinement indicator selected for p-refinement is evaluated on the resulting grids,
and refinements are performed wherever indicated.
In contrast to h-refinement,
p-refinement can be performed as a fully grid-local operation,
without the need for intermittent communication between processes.
Only once all refinement and coarsening operations have been completed,
the resulting refinement levels are communicated between neighboring grids to ensure proper allocation of boundary data buffers.

Finally, a second load balancing step is performed,
as the relative computational load of grids may have changed when their resolution was changed.

\subsubsection{Refinement Indicators}\label{sec-amr-indicators}

To determine whether a grid should be refined, coarsened, or kept at
its current level, one or more indicator functions are used.

Each indicator fulfills the following criteria:
\begin{enumerate}
\item It is grid-local, meaning it requires only data from a single grid to be evaluated.
\item It evaluates to a single real number for each grid, which can be compared against bounds set by the user.
  Values higher than a set threshold trigger refinement, while values lower than a set threshold allow for coarsening.
\item It returns a dimensionless value, so it can be used in a problem-agnostic way.
\end{enumerate}

\begin{figure}[t!]
\includegraphics[width=\columnwidth]{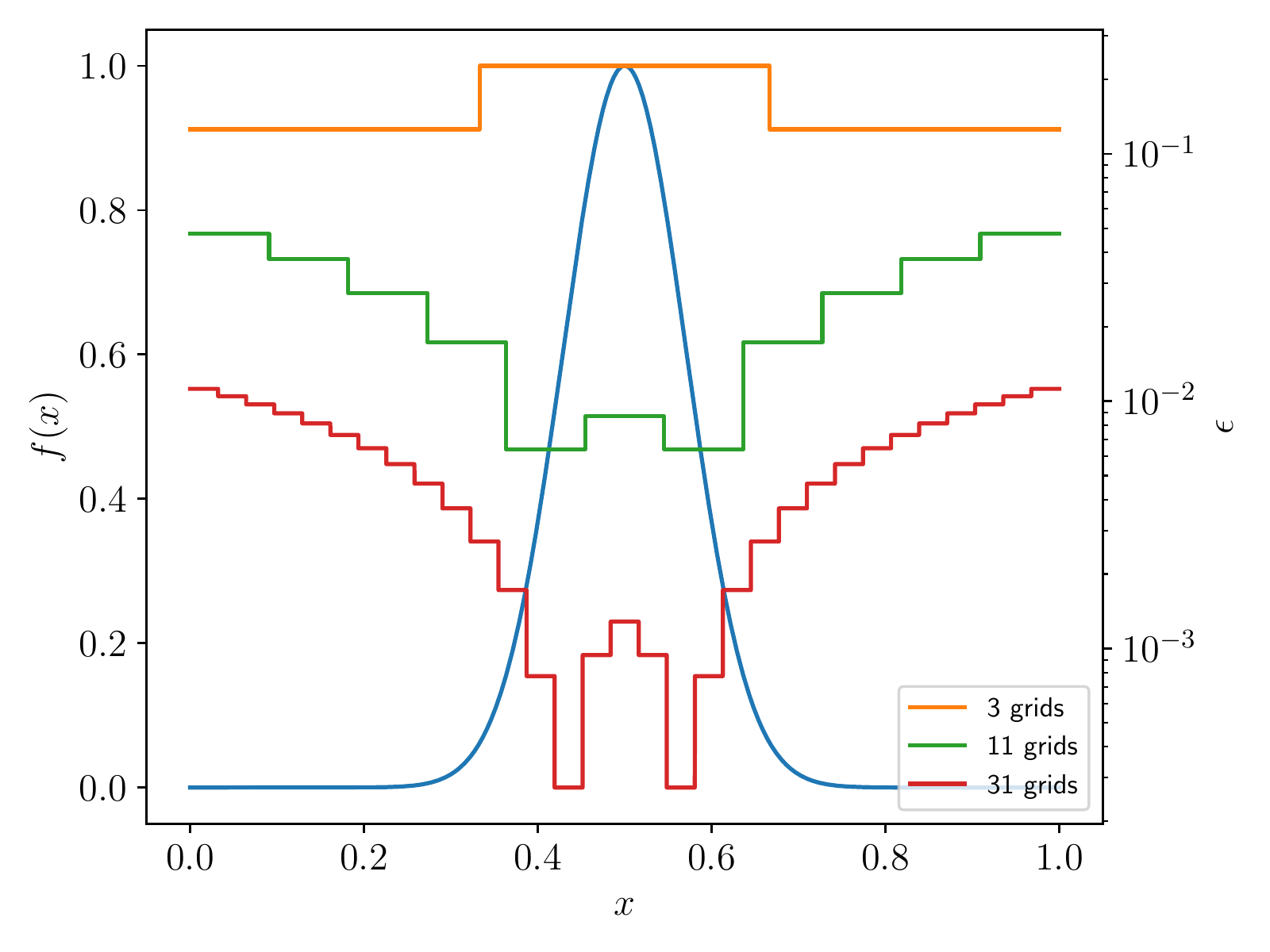}
\caption{Values returned by the smoothness estimation based refinement indicator evaluating a Gaussian (blue), at different grid sizes.}\label{fig:indicator-values-smoothness}
\end{figure}

\begin{figure}[t!]
\includegraphics[width=\columnwidth]{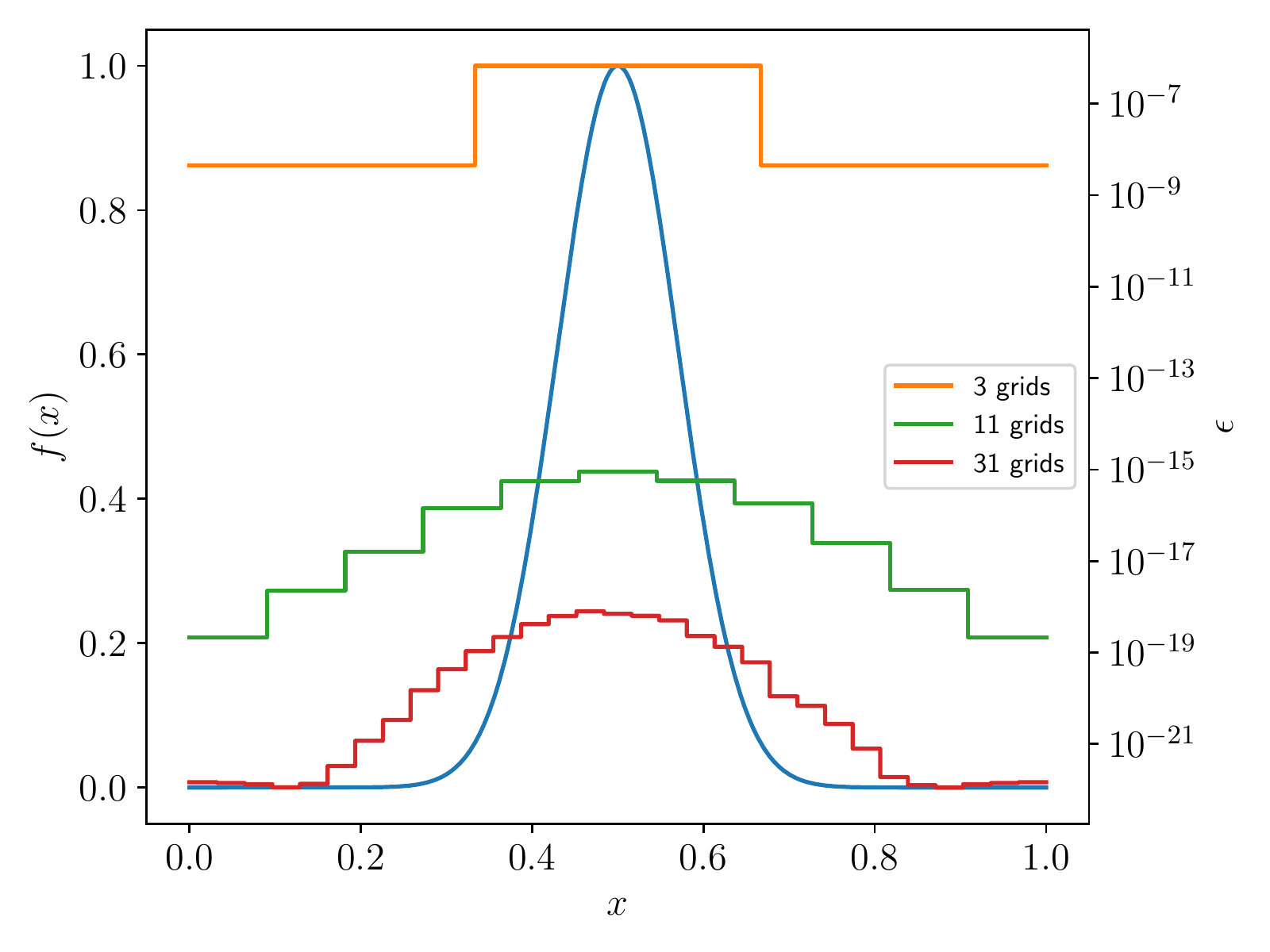}
\caption{Values returned by the truncation error based refinement indicator evaluating a Gaussian (blue), at different grid sizes.}\label{fig:indicator-values-error}
\end{figure}

Indicators can interchangably be used to drive h- or p-refinement, or both.
Both h- and p-refinement are assigned their own indicator function with their own bounds for the returned value.
See Figs.~\ref{fig:indicator-values-smoothness} and \ref{fig:indicator-values-error} for examples of different refinement indicator functions applied to a Gaussian at different h-refinement levels.

\paragraph{Truncation Error Estimate}\label{sec-error-indicator}
One tool to gauge the quality of data representation on spectral grids is to consider the decay of the coefficients of the spectral series.
For a smooth function,
when using an appropriate polynomial basis,
these coefficients will decrease exponentially as the order increases,
given enough resolution to capture high orders \cite{Boy01}.
The magnitude of the highest order coefficient can thus be used to estimate the truncation error of the series.
We use this to construct a refinement indicator similar to that described in \cite{Szi14},
which aims to keep the estimated truncation error below a specified value everywhere on the domain.

To compute this indicator,
we extract the spectral coefficients $c_i$ from the nodal representation.
For a given function basis $\{\varphi_i(x)\}$ and nodes $\{x_i\}$,
\begin{align}
  u(x) &= \sum_j c_j \varphi_j(x), \quad u_i = u(x_i), \quad V_{ij}=\varphi_j(x_i),
  \\
  u_i &= V_{ij} c_j, \quad c_i = V^{-1}_{ij} u_j . 
\end{align}
Fig.~\ref{fig:hprefexample} shows an example of the $c_i$.
For the indicator, we first compute the spectral coefficients $c_i$ along each line of grid points.
By default,
these coefficients are then normalized with respect to $c_0$,
which results in a measure of relative truncation error.
This can optionally be disabled to estimate the absolute error instead.
Some of the higher modes are effectively eliminated due to filtering,
their coefficients are therefore removed from consideration.
For the remaining coefficients, we construct a single sequence 
\begin{align}
  \tilde{c}_i = \sqrt{ \frac{1}{N} \sum_{k = 1}^{N}{c_i^k} } \quad\quad i = 1,2,\dots,\tilde{n}
\end{align}
of the root mean square of the $i$-th coefficient,
where $\tilde{n}$ is the highest non-filtered mode,
and $k$ enumerates all $N$ lines of grid points,
so $N = d \cdot n^{d-1}$ for a $d$-dimensional grid of $n^d$ points.

We then fit a simple model of exponential decay to the $\tilde{c}_i$,
and evaluate the resulting function at the highest non-filtered order to obtain the final indicator value
\begin{align}
  \varepsilon &= 10^{a \tilde{n} + b} \,,
\end{align}
where $a$ and $b$ are the slope and offset as obtained by a linear least squares fit on the logarithm of the $\tilde{c}_i$.
This $\varepsilon$ is returned as the indicator value.

Because the exponential decay of coefficients only sets in at sufficiently high order,
and accumulated round-off errors due to finite machine precision prevent the accurate computation of coefficients for very high orders,
resulting in a `roundoff plateau',
this method overall tends to underestimate the slope of the decay,
thus overestimating the total truncation error.

This type of indicator is naturally suited for driving p-refinement,
since it directly corresponds to the success or failure of a series of particular polynomial order to represent the data.

\paragraph{Smoothness Estimate}\label{sec-smoothness-indicator}

A well tested heuristic for determining the need for mesh refinement in an area is an estimate of the form

\begin{align}
  \epsilon &= \sqrt{\frac{1}{N} \sum\limits_{i}^{N} \frac{
      \sum\limits_{k,l} \left( \diff.sr.{u}{x_k,x_l}[x_i] \right)^2
    }{
      \sum\limits_{k,l} \left( \frac{\left|\diffp{u}{x_k}\right|_L + \left|\diffp{u}{x_k}\right|_R}{\Delta_l} + \varepsilon \left|\diffp{}{x_k,x_l}\right| \left|u\right| \Big|_{x_i} \right)^2
    }
  } \,,
\end{align}
similar to the indicator originally described in \cite{Loe87},
and adapted to spectral grids.

Here, $\left|\diffp{u}{x_k}\right|_L$ and $\left|\diffp{u}{x_k}\right|_R$ refer to the first derivatives of the solution at the left and right boundaries of the spectral element, respectively,
and $\Delta_l$ is the size of the element along the $l$-th dimension.
The term following $\varepsilon$ is computed by taking the absolute value of the derivative matrix $D_{kl} = D_k \cdot D_l$ elementwise,
and applying it to the piece of the state vector containing the variable $u$,
also taking the absolute value elementwise.
Effectively, we compute a normalized version of the second derivatives
on the grid, where the normalization is based on an upper bound on
first derivatives. If the first derivatives are small, then the term
proportional to $\varepsilon$ provides an alternative
normalization. It acts as a filter to prevent small high frequency
`ripples' from triggering unwanted, and potentially cascading,
refinement.

This type of indicator originates in finite element methods using linear elements,
where the magnitude of second derivatives is justified as an error estimate.
Because a spectral element of order $n>2$ can still represent 2nd order polynomials exactly,
it is less obvious why the indicator would give meaningful results.
In practice, however, using it as a heuristic leads to refinement in exactly those regions that are `non-smooth',
as well as the regions where the solution shows the strongest features.

It is a natural choice as the indicator used by the h-refinement portion of the algorithm,
as it pushes the algorithm to subdivide grids until each represents an approximately linear piece of the solution.

\begin{figure}[t!]
\includegraphics[width=0.7\columnwidth]{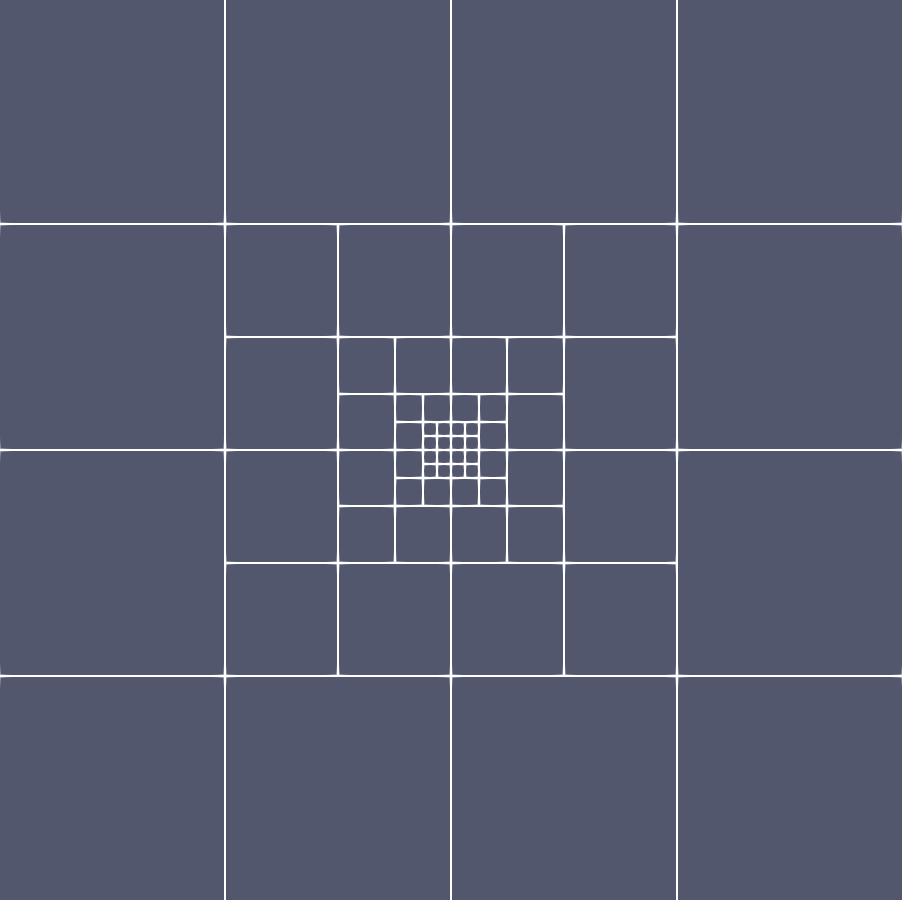}
\caption{A grid structure resulting from the use of a static refinement indicator such as (\ref{eq:center-refinement}).}\label{fig:center-refinement}
\end{figure}

\paragraph{Static Indicators}\label{sec-static-indicators}
Instead of using the data on a grid to determine its refinement status,
it is also possible to construct indicators that result in a static,
yet heterogeneous grid structure,
for example a domain with high resolution near one or more defined centers,
and progressively lower resolution further away from them.
Such a scheme can be described in terms of a target level $l$ which depends on the distance $d$ from the closest center,
for example
\begin{align}
  l &= \lfloor \log_2\left( \frac{a}{d} \right) \rfloor\,,\label{eq:center-refinement}
\end{align}
where $a$ determines the size of the refined region.
Schemes such as this,
which directly result in a target level,
can easily be made to fit the above paradigm of returning a real number to be compared to a set interval,
by returning for example $+1$ if the current level is below the target level,
$-1$ is the current level is above it,
and $0$ if the current level matches the target level.
Setting the accepted interval to $[-0.5, 0.5]$ then results in the desired refinement operations being applied
(see Fig.~\ref{fig:center-refinement}) for an example of the resulting grid structure).
Applying this type of refinement indicator in combination with for example a center-of-mass detection will result in an AMR scheme that guarantees that regions of physical interest,
such as orbiting compact objects,
are always covered by highly resolved mesh regions that dynamically follow them.

\subsubsection{Load Balancing}\label{sec-load-balancing}

In order to ensure an even distribution of computational work across
the available CPU cores, grids are shifted between processes during
AMR operations.  Since each boundary shared by two or more processes
necessitates data exchange between these processes, we also seek to
minimize the amount of such boundaries, or maximize data locality.

\begin{figure}[t!]
\includegraphics[width=0.7\columnwidth]{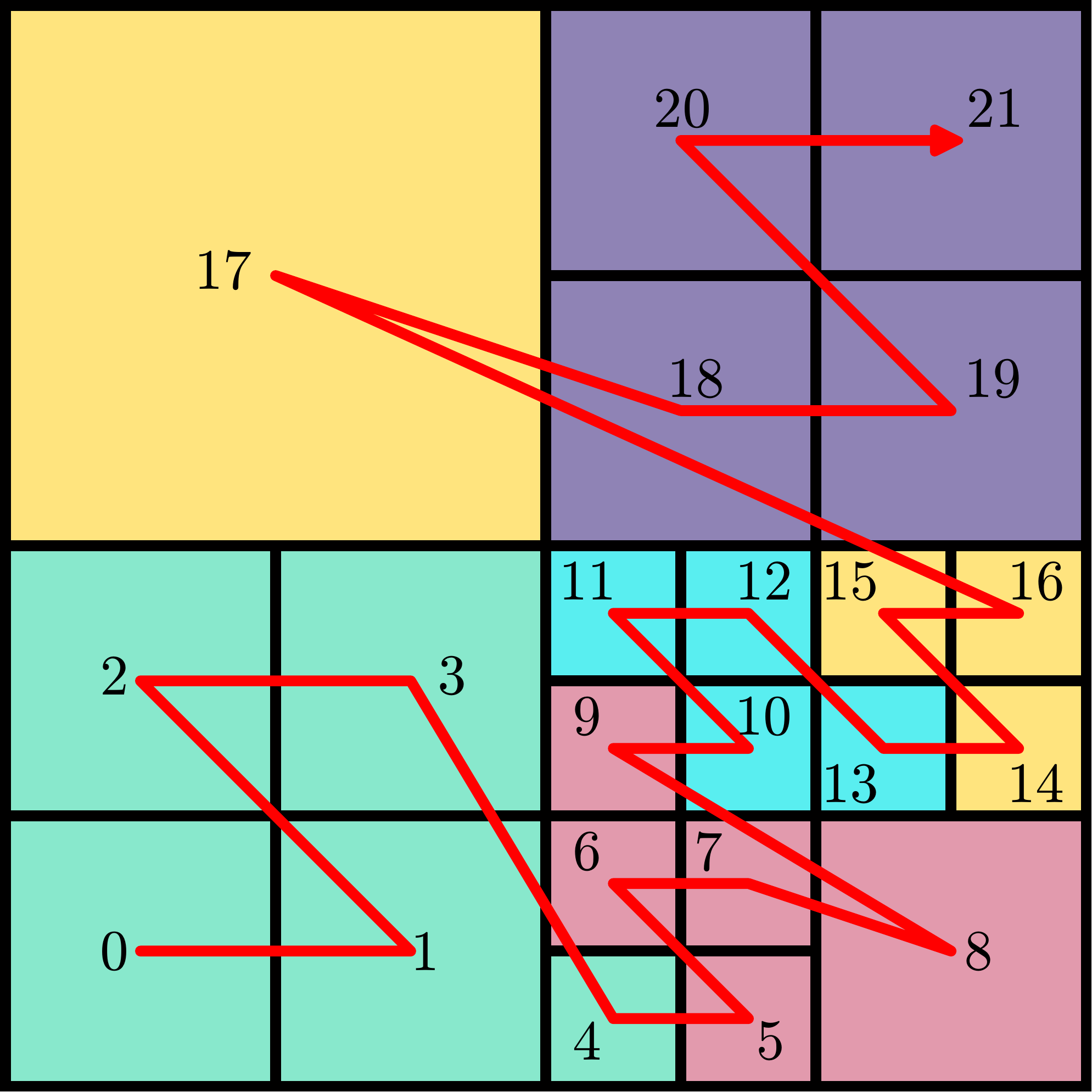}
\caption{An example of a grid configuration with the z-order curve determining the internal ordering of grids. Different colors show a possible division of 22 grids among 5 processes, demonstrating approximate data locality.}\label{fig:z-curve}
\end{figure}

This is achieved by arranging all grids into a list in order of their intersection with a space-filling z-order curve,
also known as a Morton curve~\cite{Mor66},
and then partitioning this list into as many sections as processes are used
(see Fig.~\ref{fig:z-curve}).
Similar partitioning schemes are used in other codes,
for example~\cite{DasZapCoo21} also uses a Morton curve for domain partitioning.
See also~\cite{Bad12} for a comprehensive treatment of space filling curves and their numerical applications.
We specifically use a z-order curve over the more common Hilbert curve
in order to simplify the application of grid index offsets for neighbor tracking
during refinement operations.
We find that this simple mechanism fulfills both of the goals set out
for load balancing.

As mesh refinement operations modify the grid structure (see sec.~\ref{sec-amr}),
the ordering of grids in this list is maintained such that it always corresponds to a traversal of the domain along the z-order curve.

\begin{figure}[t!]
\centering
\includegraphics[width=\columnwidth]{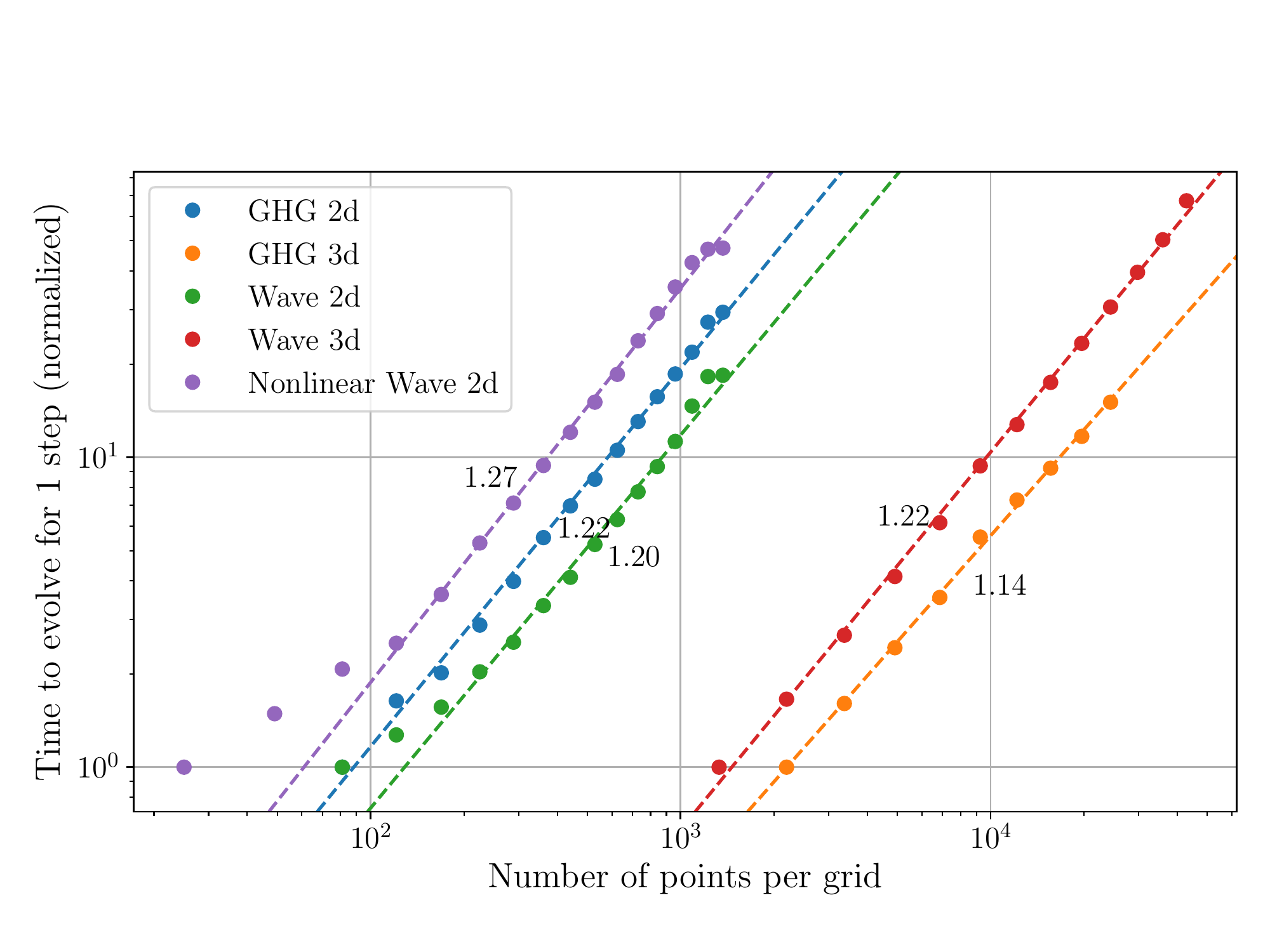}
\caption{Scaling of the workload associated with evolving a grid for different numbers of points per grid.}\label{fig:p-weights}
\end{figure}

In numerical experiments we empirically find that the runtime necessary to evolve a grid scales as
\begin{align}
  t &= n^w\,\label{eq:grid-weight},
\end{align}
where $n$ is the total number of grid points,
with powers $w$ between $1.14$ and $1.27$,
depending on the equations being evolved,
as well as the number of dimensions (see Fig.~\ref{fig:p-weights}).

\begin{table}[h]
  \centering
  \caption{Runtime scaling behaviour based on the number of gridpoints.}\label{tab:p-weights}
  \begin{tabular}{l c}
    \toprule
    Type & $w$ \\
    \midrule
    Wave eq.\ (2D) & $1.20$ \\
    Wave eq.\ (3D) & $1.22$ \\
    Nonlin.\ wave eq.\ (2D) & $1.27$ \\
    GHG (2D) & $1.22$ \\
    GHG (3D) & $1.14$ \\
    \bottomrule
  \end{tabular}
\end{table}

Consequently, each grid is assigned a weight, according to
\begin{align}
  \rho_i &= (n_i)^w \,,
\end{align}
where $n_i$ is the total number of grid points on the $i$-th grid,
and $w$ is chosen appropriately for the equations being solved.
The partitioning of the grid list is then done such that each segment contains grids with approximately the same total weight.
It should be noted that this represents a simplified model which combines both the amount of pointwise arithmetic,
which depends on the particular system being evolved,
and independent per-grid overhead (for example computation of derivatives) depending on the particular equations being solved, into a single empirical
parameter, $w$.

\subsubsection{Boundary Data Exchange}\label{sec-boundary-exchange}
The pseudospectral method used by \texttt{bamps} requires the exchange of data on the internal boundaries between grids.
The optimal way to accomplish this can and will depend on the particular MPI implementation used.
We use the following algorithm,
applied in synchronous phases accross all MPI processes.

\begin{enumerate}
\item Determine how many boundaries the MPI process shares with each other MPI process.

\item For each MPI process with shared boundaries, generate a list of identifiers $t$,
from which the details of the necessary MPI message can be uniquely determined,
\begin{flalign}
  \begin{split}
    t =&\ 24 \cdot n_g + 4 \cdot n_{dir} + n_{ne}\,,
  \end{split}
\end{flalign}
where $n_g$ is the index of the sending grid,
$n_{dir}$ is a number between 0 and 5 which encodes a direction,
and $n_{ne}$ is a number between 0 and 3 which specifies which of up to 4 neighboring grids information is sent to.
This creates a pair of identifiers for each shared boundary,
one for sending the local data,
and one for receiving data from a remote MPI process.
These identifier pairs are stored together as a single entry in the list of messages,
which can be sorted by either one.

\item Sort all lists of communications with MPI processes of higher rank by the send-identifier,
  and sort the other lists by the receive-identifier.
  This ensures that MPI processes sharing several boundaries have identical lists of communication identifiers,
  in the same order.

\item Sort the list of lists by length, to ensure the shortest ones are handled first.
  We find that this significantly reduces the time other processes spend waiting on communications to be initiated, in some cases by up to 40\%.

\item For each list of communications, initiate both sending and receiving operations as asynchronous MPI calls,
  using the list index as the MPI message tag.

\item While MPI communications are ongoing, perform all boundary data exchanges that are entirely local to each MPI process.

\item Wait for all MPI communications to finish.

\end{enumerate}

Once all boundary information has been transferred to the neighboring grids,
this data is interpolated to match the resolution of the receiving grid.

For $n_p$ different grid resolutions accessible by p-refinement,
there are $n_p^2$ possible combinations between equally sized grids,
each requiring a different interpolation matrix.
Since grids may also share boundaries with other grids either half or twice their size,
which may overlap either in the upper or lower half of their respective extents,
this number is further multiplied to give a total of $5 n_p^2$ possible cases.
In practice,
only a small subset of these cases will be reached during any given simulation.
Therefore, we do not compute every possible interpolation matrix in advance,
and instead generate them on demand.
We then utilize memoization (caching) to minimize duplication of work,
that is each MPI process keeps a cache of previously required interpolation matrices,
to be re-used if the same case appears again later.
Each interpolation matrix that is required is thus only generated once on each process where it is needed,
and unneeded matrices are never generated.

\subsection{Performance and Scaling}\label{sec:performance}

To evaluate the performance and scaling behavior of \texttt{bamps},
a series of benchmark runs was performed on a varying number of CPU cores.
For these runs, static grid configurations were chosen,
in order to control the amount of work per CPU core.
Each configuration was run 3 times, and the measured times were averaged.
The times themselves were measured using built-in timers,
capable of profiling specific sections of the code,
including the runtime of the entire \texttt{main} routine.
With this, both strong and weak scaling tests were performed.
We use an axisymmetric subcritical Brill wave collapse simulation as our test case,
see Sec.~\ref{sec:brillwave} for details on the evolution system.

\begin{figure}[t!]
\includegraphics[width=\columnwidth]{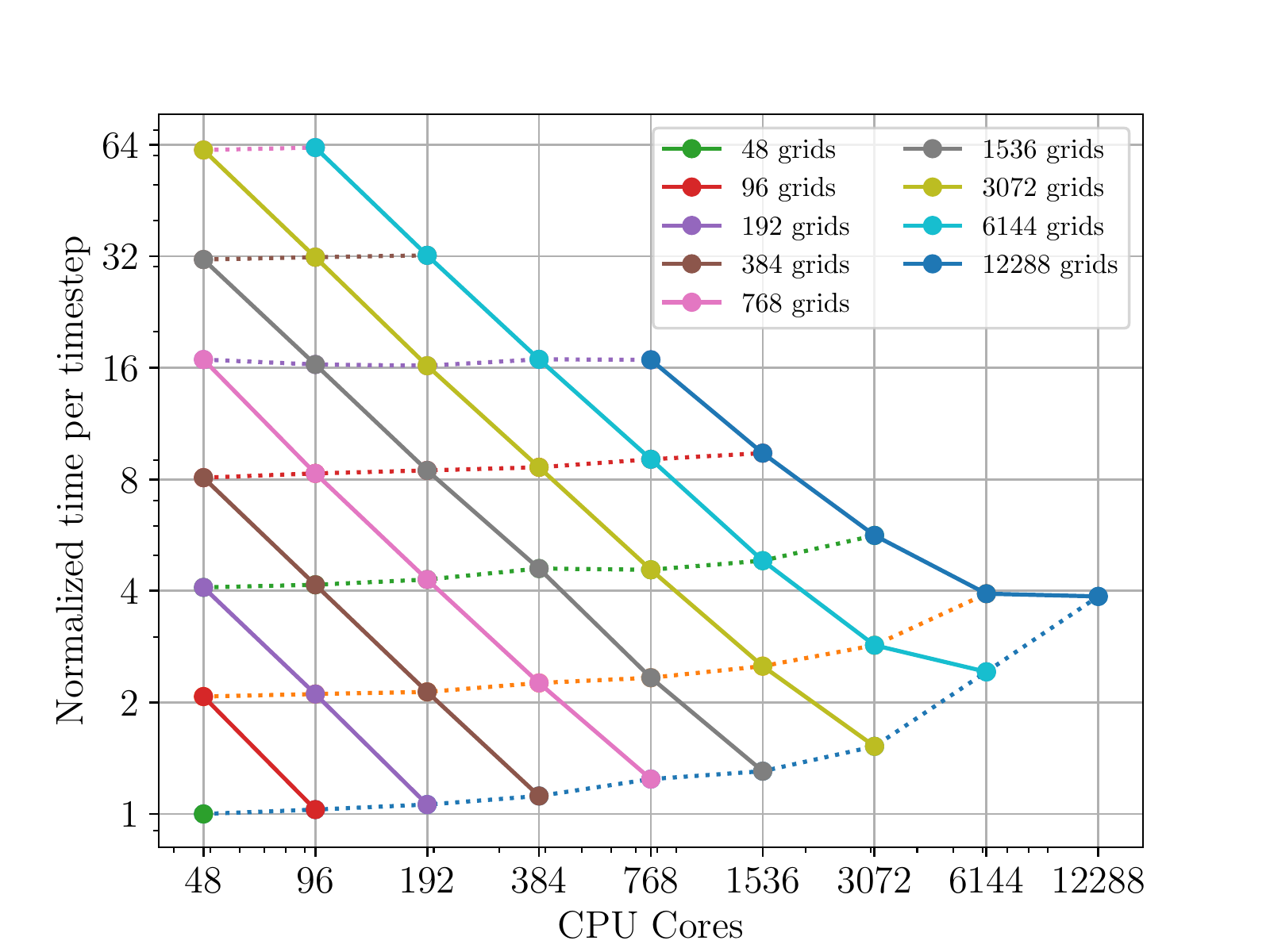}
\caption{Strong and weak scaling of \texttt{bamps} for different static grid configurations,
  based on benchmarks performed on SuperMUC-NG,
  evolving the GHG system in 2d.
  Solid lines represent a constant total number of grids,
  dotted lines represent a constant number of grids per CPU core.}
  \label{fig-scaling-grid}
\end{figure}

\begin{figure}[t!]
  \includegraphics[width=\columnwidth]{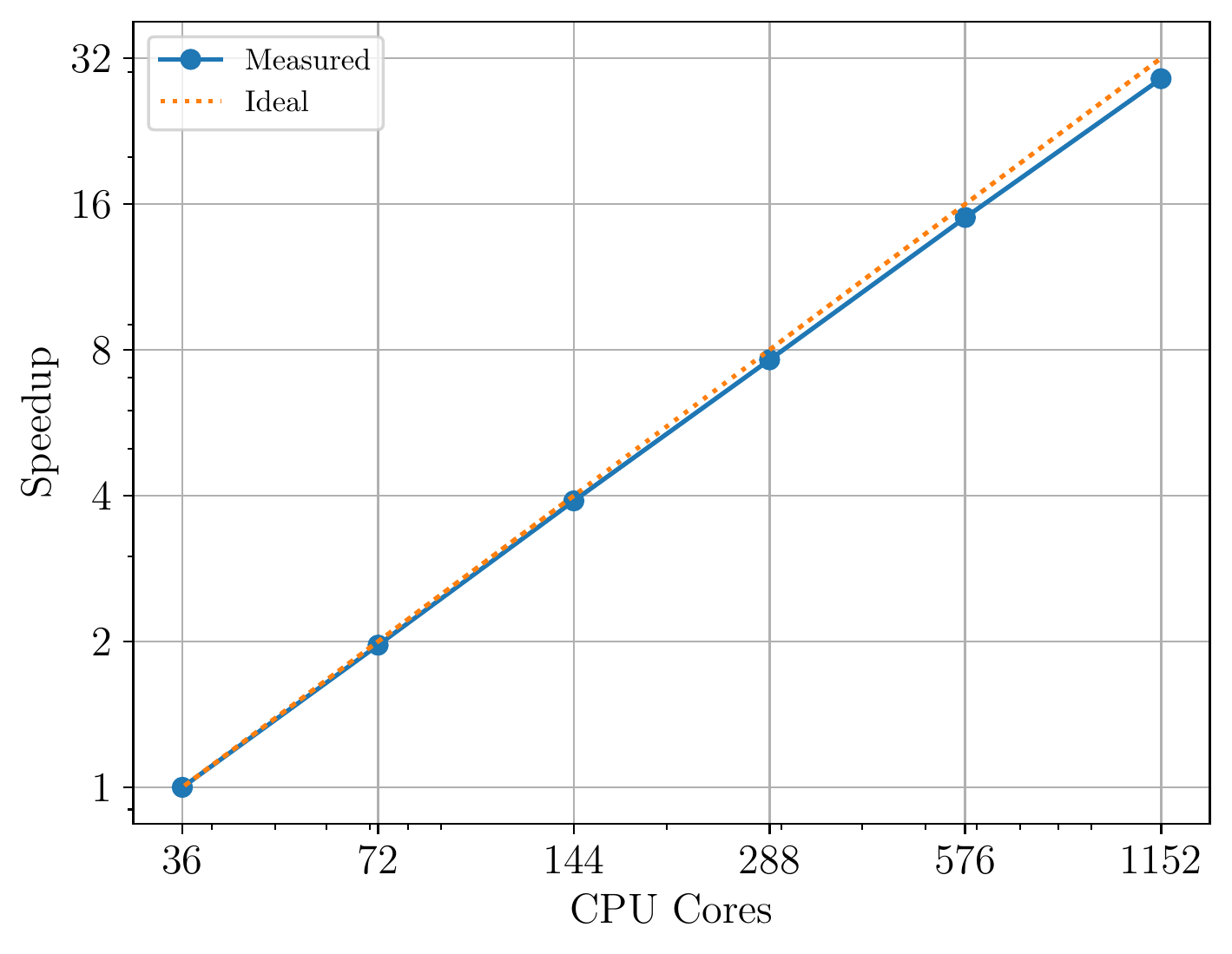}
  \caption{Strong scaling of \texttt{bamps} for a static grid configuration of 9216 grids.}
  \label{fig-strong-scaling}
\end{figure}

Strong scaling refers to the performance increase, as measured by the
lower runtime (speedup), when distributing the same amount of work
over more CPU cores.  Ideal strong scaling would be achieved if a
doubling of the number of CPU cores resulted in halving the
necessary time for the same simulation.  In practice, most programs
have a non-parallelizable part, which leads to their speedup following
Amdahl's law, with diminishing returns for increasing numbers of CPU
cores.

\begin{figure}[t!]
  \includegraphics[width=\columnwidth]{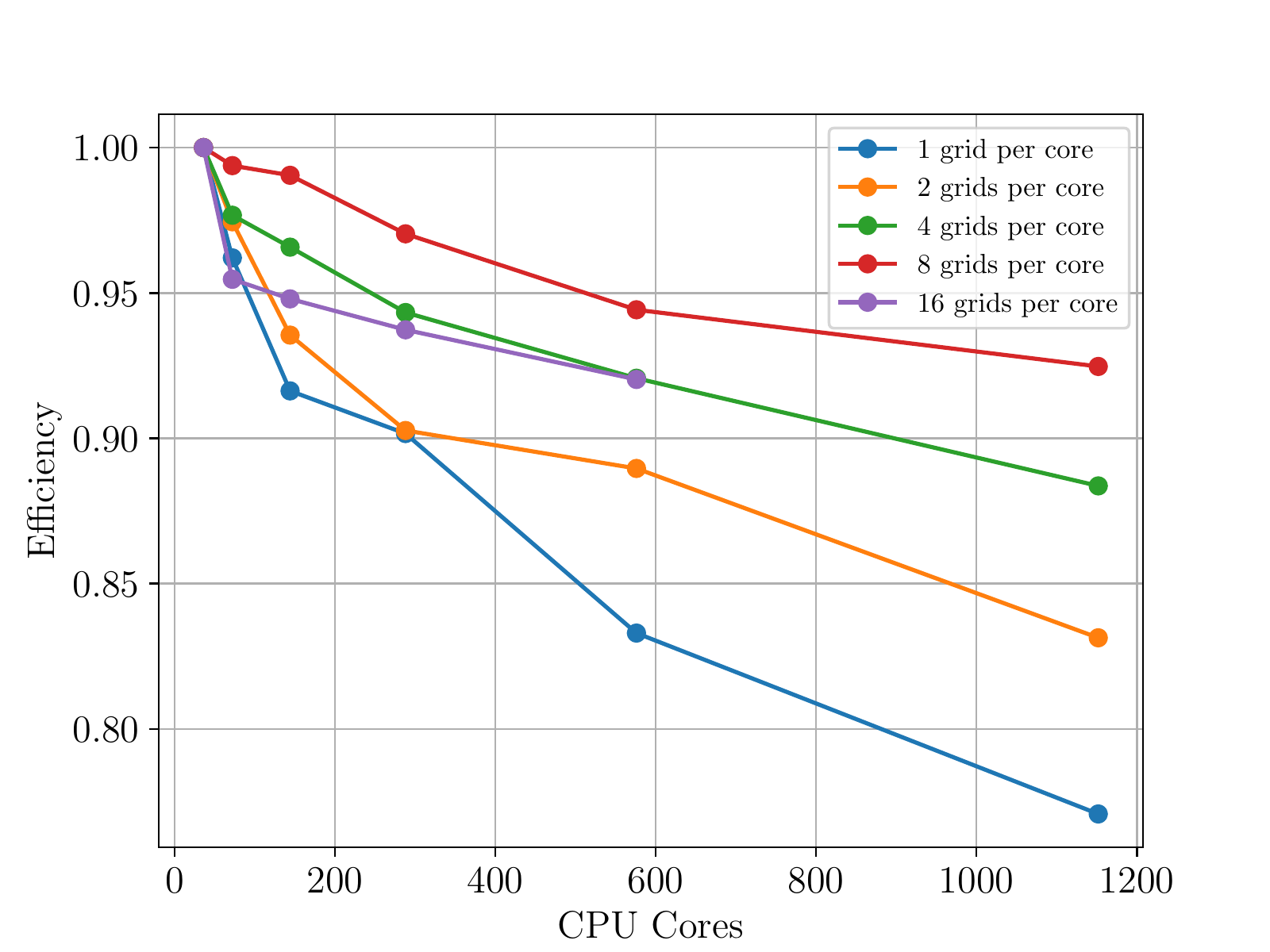}
  \caption{Weak scaling efficiency of \texttt{bamps} for a static grid configuration.}
  \label{fig-weak-scaling}
\end{figure}

Weak scaling refers to the ability to solve larger problems efficiently
when provided with more CPU cores.
It is measured by increasing both the problem size and the number of CPU cores by the same factor,
and observing the change in computation time.
Ideal weak scaling would be achieved if the required time remains the same under this change.
The weak scaling is often measured by the weak scaling efficiency,
determined by

\begin{flalign}
  e &= \frac{t_0}{t} \cdot \frac{n}{n_0}\,,
\end{flalign}
where $t$ is the total computation time summed over all CPU cores,
and $n$ is the number of utilized CPU cores.
$t_0$ and $n_0$ are those quantities for a selected (small) reference run.

In Fig.~\ref{fig-scaling-grid}, both strong and weak scaling of \texttt{bamps} are shown.
The solid lines represent series of runs showing strong scaling,
as the computation time required decreases at the same rate as the number of CPU cores is increased.
The dotted lines show a series of runs demonstrating weak scaling,
as the amount of work per CPU core remains constant along them,
and the amount of computation time required also remains almost constant.

The strong scaling behaviour of \texttt{bamps} is also shown separately in Fig.~\ref{fig-strong-scaling},
and Fig.~\ref{fig-weak-scaling} shows the weak scaling efficiency.
Provided the workload of a single CPU core is chosen to be sufficiently large
(8 grids per CPU core for the GHG simulations studied here),
we consistently observe a weak scaling efficiency above 90\%.

With AMR enabled, the scaling behaviour is expected to be slightly worse,
since the refinement algorithm involves global communication between all MPI processes during the load balancing procedure.
However, because the AMR algorithm is only invoked every 100 timesteps
(or even more rarely),
we do not expect it to have a noticable impact on the overall scaling behaviour.

Given the complexity of hp-refinement even for fixed refinements, in this particular example the scaling of \texttt{bamps} is excellent up to 1000 CPU cores and satifactory up to 6000 CPU cores. These numbers change when considering runs with more grids, and in particular when moving from 2d to 3d spatial grids.

Overall, the strategy to consider MPI parallelization with single
grids defining data granularity is successful for a wide range of
configurations, down to even a few 2d grids per process. In part this is
a feature of the complexity of the Einstein equations, since the
amount of data and work per grid tends to be large compared to the
parallelization overhead. In general, this is also expected for
spectral element methods, if only face-local data is exchanged at
element interfaces. More fine-grained algorithms are available, for
example task-based parallelism as employed in
\cite{KidFieFou16,DasZapCoo21}, but for the present applications of
\texttt{bamps} the single-grid granularity performs well.

\section{Non-Linear Wave Equation Model}\label{sec:nonlinear-wave-equation}

To evaluate the capability of the AMR system in resolving strongly varying data during an evolution,
we consider the nonlinear wave equation
\begin{align}
  \square \psi + A_1 \nabla_a\nabla^a \psi &= 0\,,
\end{align}
which corresponds to `model 1' in \cite{SuaVicHil21},
choosing $A_1 = 1$.
We apply a first-order reduction using the reduction variables $\Pi = \partial_t \psi$ and $\phi_i = \partial_i \psi$,
resulting in the system
\begin{align}
  \partial_t \psi &= -\Pi \,,\\
  \partial_t \phi_i &= -\partial_i \Pi + \gamma_2 \partial_i \psi - \gamma_2 \phi_i \,,\\
  \partial_t \Pi &= -\partial_i \phi^i - A_1 (\phi_i \phi^i - \Pi^2) \,.
\end{align}

Solutions to this equation can be built analytically,
by first constructing a solution to the linear equation $\square~\psi~=~0$ from partial waves as
\begin{align}
  \varphi &= \sum\limits_{l=0}^{\infty} \sum\limits_{m={-l}}^{l} \varphi_{lm}(t, r) Y_l^m(\theta^A)\,,
\end{align}
where $(r,\theta^A)$ are the usual spherical coordinates,
and $Y_{lm}(\theta^A)$ the spherical harmonics.
Applying the deformation function
\begin{align}
  D(\varphi) &= A_1^{-1} \log(1+ A_1 \varphi)
\end{align}
then yields the solution $\psi = D(\varphi)$.

For this comparison, we evolve data built from a pure $(l=2, m=0)$-wave,
such that
\begin{align}
  \begin{split}
  \varphi(t, r, \vartheta) = \frac{1}{4}\sqrt{\frac{5}{\pi}} \bigg( &\frac{3}{r^3} \left[ F(t_{-}) - F(t_{+}) \right] \\
           + &\frac{3}{r^2} \left[ F^\prime(t_{-}) + F^\prime(t_{+}) \right] \\
           + &\frac{1}{r} \left[ F^{\prime\prime}(t_{-}) - F^{\prime\prime}(t_{+}) \right] \bigg) \left(3 \cos^2 \vartheta - 1\right)\,,
  \end{split}
\end{align}
where $t_{-} = t-r$ and $t_{+} = t+r$ are the retarded and advanced time,
respectively
(see \cite{GunPriPul93} for details on this construction).

For the seed function $F(t)$ we choose an offset Gaussian
\begin{align}
  F(t) &= A\,\mathrm{e}^{-(t+1)^2} \,.
\end{align}
For sufficiently high amplitudes this solution is known to `blow up' after finite time~\cite{SuaVicHil21},
while for smaller amplitudes the solution continues to exist.
We refer to these cases as supercritical and subcritical, respectively.
In order to obtain strong features in the solution,
we choose an amplitude of $A = 1.6784366869120966$,
which we find to be barely subcritical when evolved with the highest resolution used here.

\begin{figure}[t!]
\includegraphics[width=\columnwidth]{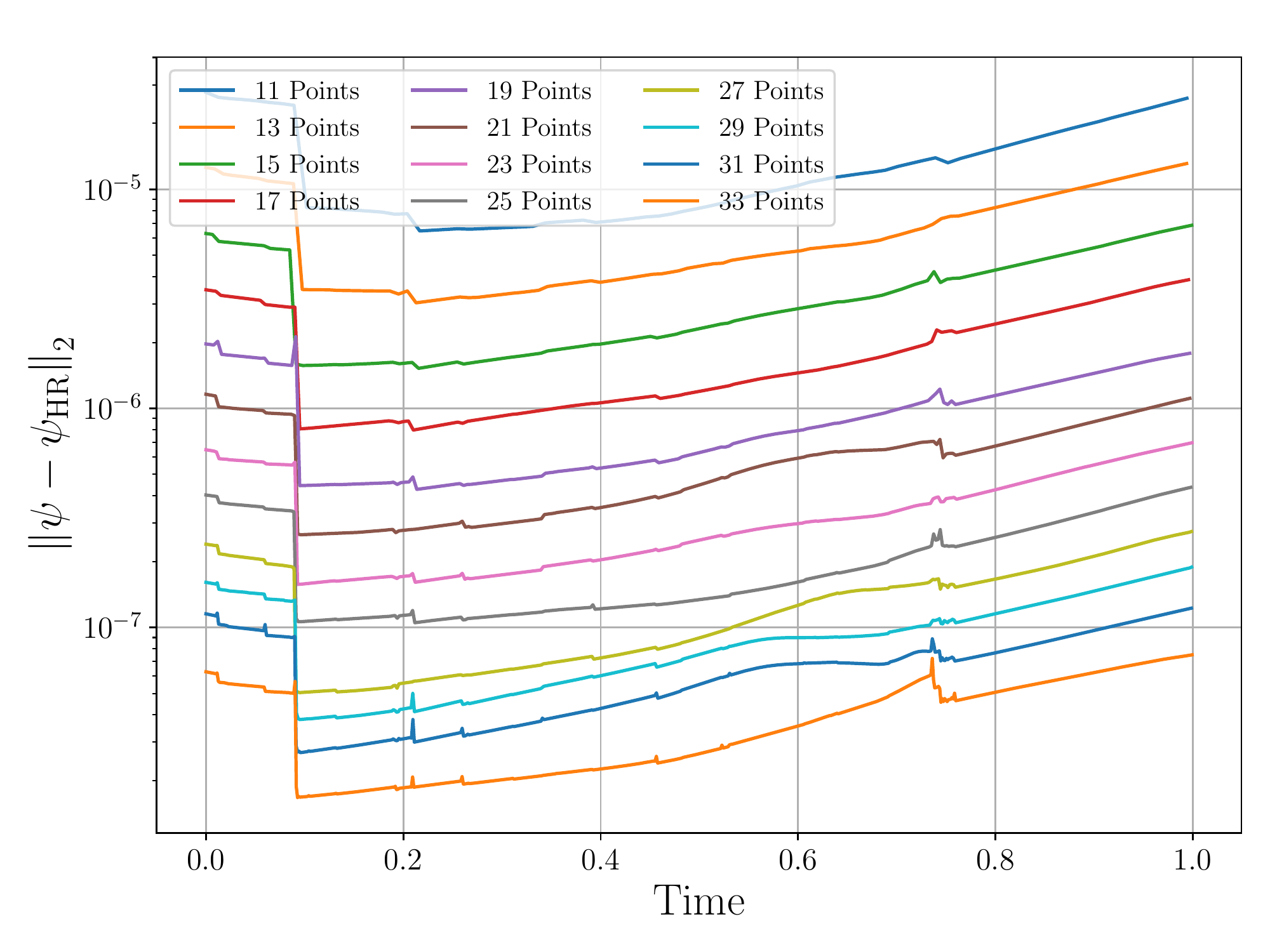}
\caption{Non-linear wave model, convergence of the numerical solution at different per-grid resolutions, as compared to a reference simulation with $35\times35$ points per grid. For the sake of comparison, the same h-refinement schedule was applied to all runs.}\label{fig:nonlinear-wave-convergence}
\end{figure}

For a systematic convergence test,
we employ the concept of a refinement schedule.
A run is performed for a specific choice of hp-refinement parameters
(typically for the lowest feasible resolution),
and the time-dependent sequence of h-refinements is recorded.
Subsequent runs can use the same `h-refinement schedule' while varying the p-refinement.
Using this technique,
we find the expected exponential convergence of the numerical solution as grid points are added,
which corresponds to adding terms to the spectral series (see Fig.~\ref{fig:nonlinear-wave-convergence}).

To determine the impact of AMR on the accuracy on the time evolution,
we evolve this system using 
\begin{enumerate*}[(i)]
\item AMR, \label{item-amr}
\end{enumerate*}
as well as static grid configurations corresponding to
\begin{enumerate}[(i)]
\setcounter{enumi}{1}
\item the lowest resolution accessible to AMR, \label{item-lowres}
\item the highest resolution accessible to AMR, \label{item-highres}
\item the average resource use of the AMR configuration. \label{item-avg}
\end{enumerate}

To determine an appropriate average resolution for configuration~\ref{item-avg},
the grid structure generated by AMR was analyzed post hoc,
and the total number of points used at any given timestep was averaged over the whole evolution.
A static, approximately uniform grid configuration using a similar amount of points was then found by a brute force search over all possible grid configurations.

We find that the use of AMR results in a lower total numerical error by up to three orders of magnitude,
compared to using a similar number of grid points spread uniformly over the domain.
Perhaps surprisingly,
the evolution using configuration~\ref{item-highres} shows by far the largest numerical error after the solution forms strong features,
several order of magnitude higher than even configuration~\ref{item-lowres}.
This is likely due to large amounts of round-off error piling up due to both a very large number of timesteps necessary to satisfy the Courant–Friedrichs–Lewy condition~\cite{CouFriLew28},
and grid-wise operations such as derivative computation requiring multiplication with very large matrices.
This effect is amplified in regions with large absolute values of the solution,
decreasing the absolute numerical precision.
We also observe the formation of non-physical features caused by large amounts of noise at the boundaries between grids.
This suggests that spacetime configurations with strong, but highly localized features,
as often encountered close to criticality,
are only accessible using AMR,
as neither low nor high static resolutions are capable of resolving them accurately.

\begin{figure}[t!]
  \includegraphics[width=\columnwidth]{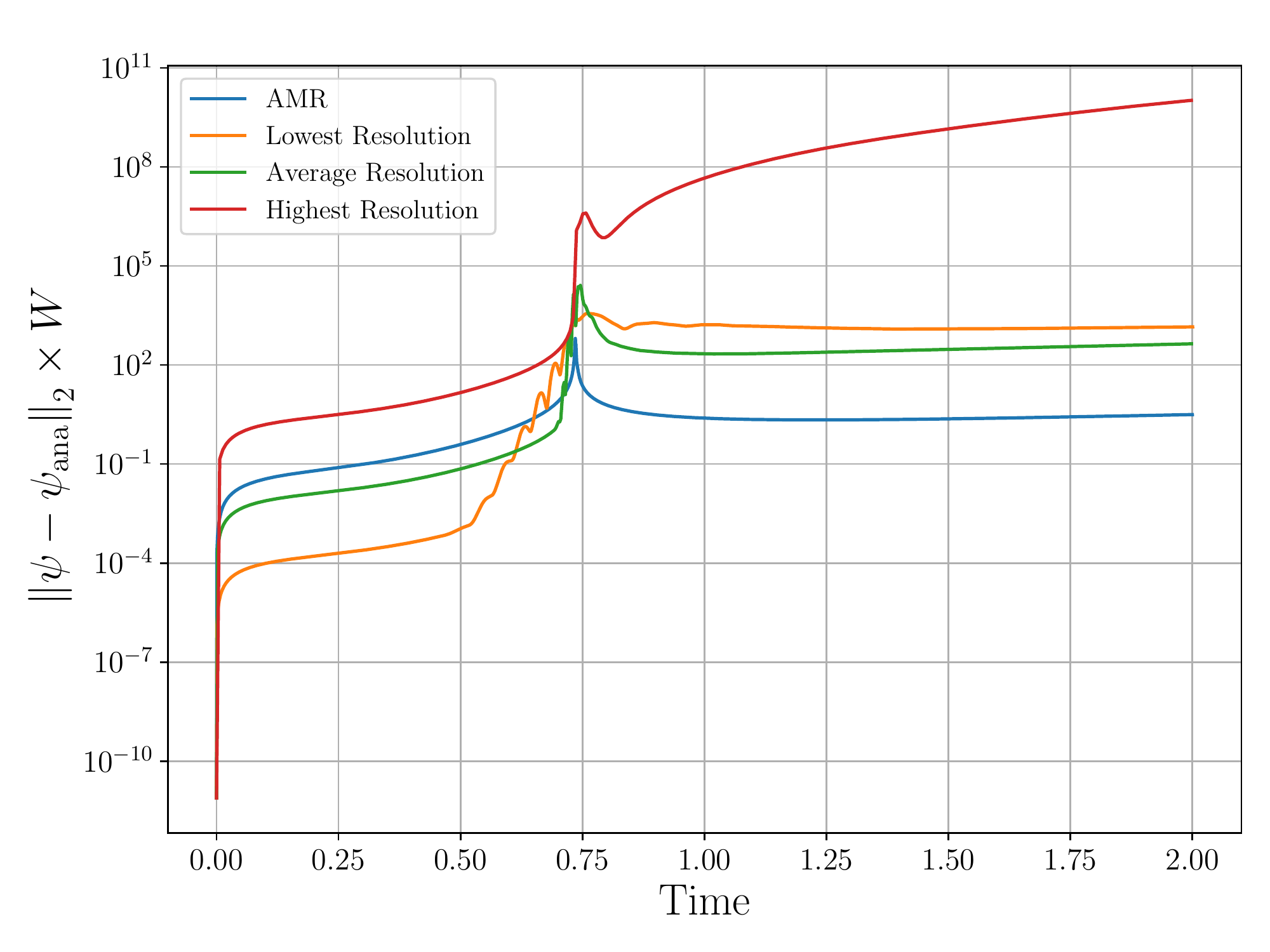}
  \caption{Efficiency for the non-linear wave model. Shown is the 
    numerical error during the evolution of a nonlinear wave equation for several static resolutions, as well as using AMR, adjusted for amount of work $W$ necessary (lower values are better).
  }
  \label{fig:nonlinear-wave-error-work}
\end{figure}

Apart from the final numerical error resulting from an evolution,
we also consider the amount of computational work necessary to evolve a given grid configuration.
Here, we use the workload (\ref{eq:grid-weight}) to compute the total work necessary to evolve data up to a given timestep as
\begin{align}
  W(k) &=  \sum\limits_{j=1}^{k} \sum\limits_{i = 1}^{N_j} (n_i)^w \,,
\end{align}
where $i$ enumerates all $N_j$ grids present at a given step $i$,
$n_i$ is the number of points on each particular grid,
and $w$ is the appropriate weighting power as shown in Tab.~\ref{tab:p-weights}.

The product of numerical error and the work required to reach it can then be used as a measure of numerical efficiency
(where a lower value corresponds to a higher efficiency).
This measure is shown in Fig.~\ref{fig:nonlinear-wave-error-work}.
The additional overhead of the AMR mechanism initially makes the evolution less efficient than a comparable but homogeneous resolution,
but once strong features form in the solution,
the additional accuracy gained via AMR leads to the full evolution being not only more accurate by several orders of magnitude,
but also more efficient in terms of work expended to obtain this accuracy.

\section{Real Scalar Field}\label{sec:scalarfield}
    
After testing AMR with a non-trivial toy model,
we test how it performs in physical scenarios of interest.
Firstly, we consider a real massless scalar field minimally coupled to the Einstein field equations,
in spherical symmetry. 
    
Denoting $g_{ab}$ the $4$d metric and $R_{ab}$ the Ricci tensor,
the Einstein equations read as
\begin{align}
  \label{eq:EFE}
  R_{ab} = 8\pi \Big(T_{ab} -\frac{1}{2}g_{ab} T\Big)\,,
\end{align}
where $T=g^{ab}T_{ab}$ and
\begin{align}
  \label{eq:energy_momentum_tensor}
  T_{ab} = \nabla_a \varphi \nabla_b \varphi - \frac{1}{2} g_{ab}(
  \nabla^c \varphi \nabla_c \varphi + m^2 \varphi^2)
\end{align}
is the energy momentum tensor corresponding to a scalar field $\varphi$.
    
We evolve initial data in time according to the $3+1$ decomposition 
\begin{align}
  \label{eq:3+1}
  ds^2 = -\alpha^2 \,dt^2 + \gamma_{ij}\,(\beta^i dt + dx^i)(\beta^j dt + dx^j)\,,
\end{align}
where $\gamma_{ij}$ is the $3$d spatial metric,
$\alpha$ is the lapse and $\beta^i$ the shift.
The normal unit vector is then $n^a=\alpha^{-1}(1, -\beta^i)$.
We write $4$d component indices with latin letters starting from $a$ and $3$d spatial ones starting from $i$.
    
The evolution equations for the matter part of the Einstein field equations follow the first order Einstein-Klein-Gordon system,
which for $m=0$ reads as
\begin{align}
  \label{eq:sf_rhs}
  \partial_t \varphi &= \alpha \pi + \beta^i \chi_i \,,\\
  \begin{split}
    \partial_t \pi &= \beta^i \partial_i \pi + \gamma^{ij} (\chi_j \partial_i \alpha + \alpha \partial_i \chi_j - \alpha \Gamma^k{}_{ij} \chi_k) \\ 
                    &\quad + \alpha \pi K + \sigma \beta^i S_i \,,
  \end{split} \\
  \partial_t \chi_k  &= \pi \partial_k \alpha + \alpha \partial_i \pi + \chi_j \partial_i \beta^j + \beta^j \partial_j \chi_i + \sigma \alpha S_i \,,
\end{align}
where $\pi$ is the time reduction variable $+ n^a \partial_a \varphi$,
$\chi_i$ is the spatial reduction variable associated to the reduction constraint $S_i := \partial_i \varphi - \chi_i$,
and $\sigma$ is a damping term.
Similarly, the metric is evolved following the generalised harmonic gauge formalism of the Einstein field equations,
  \begin{align}       
  \label{eq:ghg}
    \partial_t \tensor{g}{_a_b} &=
    \beta^i \partial_i \tensor{g}{_a_b}
      - \alpha \tensor{\Pi}{_a_b}
      + \gamma_1 \beta^i \tensor{C}{_i_a_b} \,,\\
    \begin{split}
      \partial_t \tensor{\Pi}{_a_b} &=
        \beta^i \partial_i \tensor{\Pi}{_a_b}
        - \alpha \gamma^{ij} \partial_i \tensor{\Phi}{_j_a_b}
        + \gamma_1 \gamma_2 \beta^i \tensor{C}{_i_a_b} \\
      &\quad + 2 \alpha g^{cd}
        \left( \gamma^{ij} \tensor{\Phi}{_i_c_a} \tensor{\Phi}{_j_d_b}
        - \tensor{\Pi}{_c_a} \tensor{\Pi}{_d_b}
        - g^{ef} \tensor{\Gamma}{_a_c_e} \tensor{\Gamma}{_b_d_f} \right) \\
      &\quad - 2 \alpha \left( \tensor{\nabla}{_(_a} \tensor{H}{_b_)}
        + \gamma_4 \tensor{\Gamma}{^c_a_b} \tensor{C}{_c}
        - \frac{1}{2} \gamma_5 \tensor{g}{_a_b} \Gamma^c \tensor{C}{_c} \right) \\
      &\quad -\frac{1}{2} \alpha n^c n^d \tensor{\Pi}{_c_d} \tensor{\Pi}{_a_b}
        - \alpha n^c \gamma^{ij} \tensor{\Pi}{_c_i} \tensor{\Phi}{_j_a_b} \\
      &\quad + \alpha \gamma_0 \left( 2 \tensor{\delta}{^c_{(a}} \tensor{n}{_{b)}}
        - \tensor{g}{_a_b} n^c \right) \tensor{C}{_c}\\
      &\quad - 16 \pi \alpha 
        \left (T_{ab} - \frac{1}{2} g_{ab} \tensor{T }{^c_c} \right) \,,
    \end{split}\\
    \begin{split}
      \partial_t \tensor{\Phi}{_i_a_b} &=
        \beta^j \partial_j \tensor{\Phi}{_i_a_b}
        - \alpha \partial_i \tensor{\Pi}{_a_b}
        + \gamma_2 \alpha \tensor{C}{_i_a_b} \\
      &\quad + \frac{1}{2} \alpha n^c n^d
        \tensor{\Phi}{_i_c_d} \tensor{\Pi}{_a_b}
        + \alpha\gamma^{jk} n^c \tensor{\Phi}{_i_j_c}
        \tensor{\Phi}{_k_a_b} \,,
    \end{split}
  \end{align}
where the evolved variables are the metric $g_{ab}$,
$\Pi_{ab}$ the time reduction variable corresponding to $+n^d \partial_d g_{ab}$,
and the spatial reduction variable $\Phi_{iab}$ associated to the reduction constraint $C_{iab}=\partial_ig_{ab}-\Phi_{iab}$.
The constraint damping parameters are $\gamma_1=-1, \gamma_0=\gamma_2=2$ and $\gamma_4=\gamma_5=0.5$.
$C_a = H_a + \Gamma_a$ is the harmonic constraint,
where $H_a$ is a gauge source function.
    
The type of scalar field initial data we use for these simulations has a Gaussian profile of the form 
\begin{equation}
  \label{eq:initial_phi_gaussian}
  \varphi = A ( e^{-(r - R_0)^2} + e^{-(r + R_0)^2} )\, ,
\end{equation}
with $R_0 = 3$ and with a vanishing gradient along the normal vector $n^a$,
\begin{equation}
  \label{eq:initial_pi_gaussian}
  n^a \nabla_a \varphi = 0\, .  
\end{equation}
A conformal decomposition of the metric allows to solve the ADM constraints via the extended conformal thin sandwich equations (XCTS)~\cite{BauSha10,Tic17}.
We consider a flat conformal spatial metric $\bar \gamma_{ij} = \delta_{ij}$,
with a vanishing time derivative $\partial_t \bar \gamma_{ij} = 0$,
as well as maximal slicing $K = 0$, $\partial_t K = 0$.
With these choices the XCTS equations constitute a set of coupled elliptic PDEs for the conformal factor $\psi$ and the gauge variables $\beta^i$ and $\alpha$.
The latter are simply solved by $\beta^i=0$ and $\alpha=1$,
thanks to the choice in~\eqref{eq:initial_pi_gaussian}.
The remaining XCTS equation to solve for $\psi$ is
\begin{equation}
  \label{eq:initial_data_pde_gaussian}
  0 = \delta^{ij} \partial_i \partial_j \psi
      + \pi \psi \delta^{ij} \partial_i \varphi \partial_j \varphi \,.
\end{equation}
We use Robin outer boundary conditions compatible with a $1/r$ decay.
The XCTS equation~\eqref{eq:initial_data_pde_gaussian} is then solved by means of the hyperbolic relaxation method~\cite{RueHilBug17} provided by \texttt{bamps}.

Similarly to sec.~\ref{sec:nonlinear-wave-equation},
we evolve this initial data using configurations~\ref{item-amr}-\ref{item-avg},
where again configuration~\ref{item-avg} is determined by analyzing the results of \ref{item-amr},
in order to find a setup using a comparable total amount of work to evolve.
As no analytic solution is known,
we use the integral of the constraint monitor $C_{\mathrm{mon}}$,
which aggregates both physical and reduction constraint violations,
as a proxy for numerical error.

\begin{figure}[t!]
\includegraphics[width=\columnwidth]{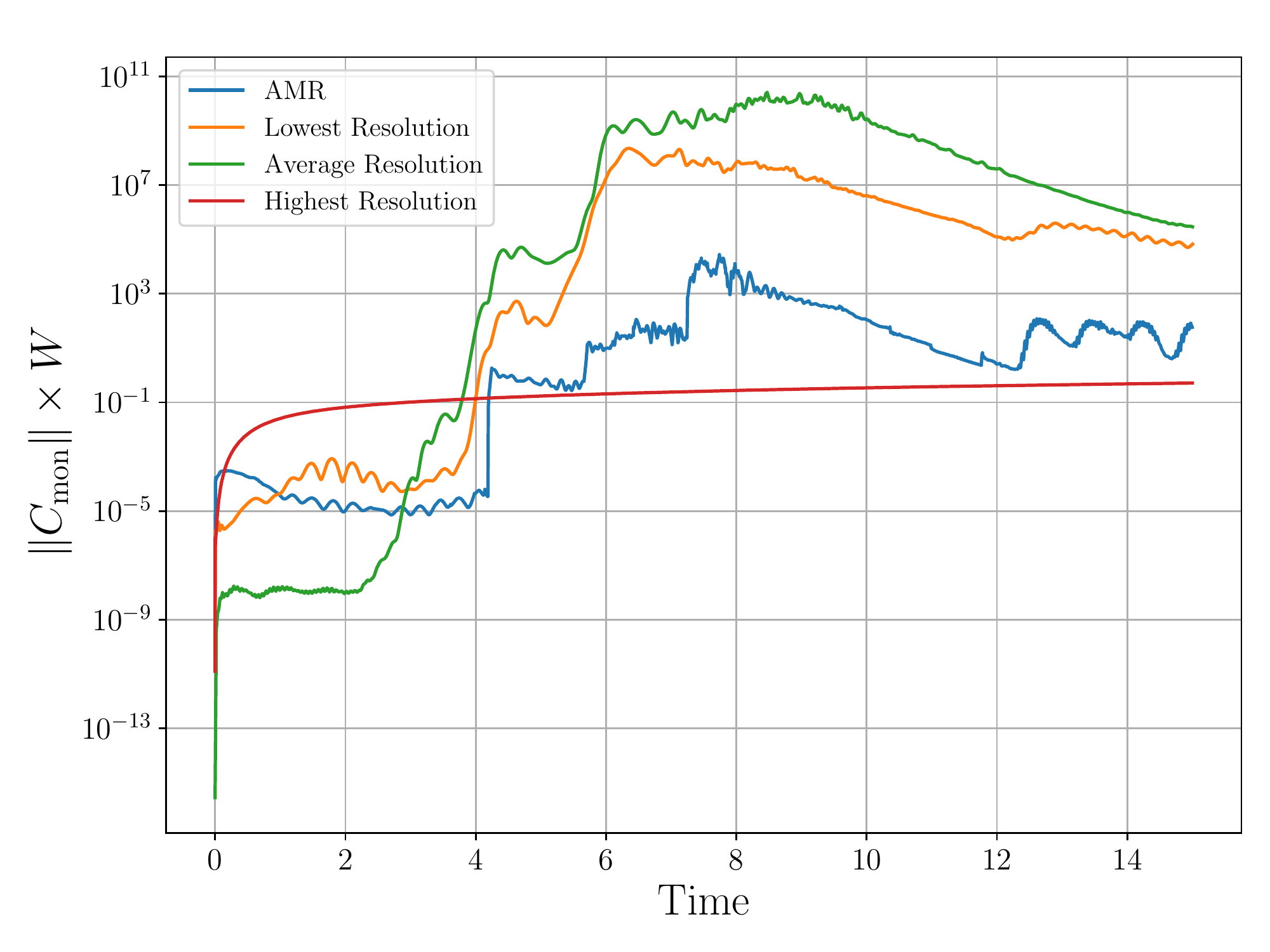}
\caption{Integral of the internal constraint violation monitor, multiplied by the cumulative workload $W$, during the time evolution of a real scalarfield using AMR, as compared to the base resolution used by the AMR, a static resolution with comparable total workload, and the highest resolution accessible to the AMR system.}\label{fig:scalarfield-error-times-work}
\end{figure}

\begin{figure}[t!]
\includegraphics[width=\columnwidth]{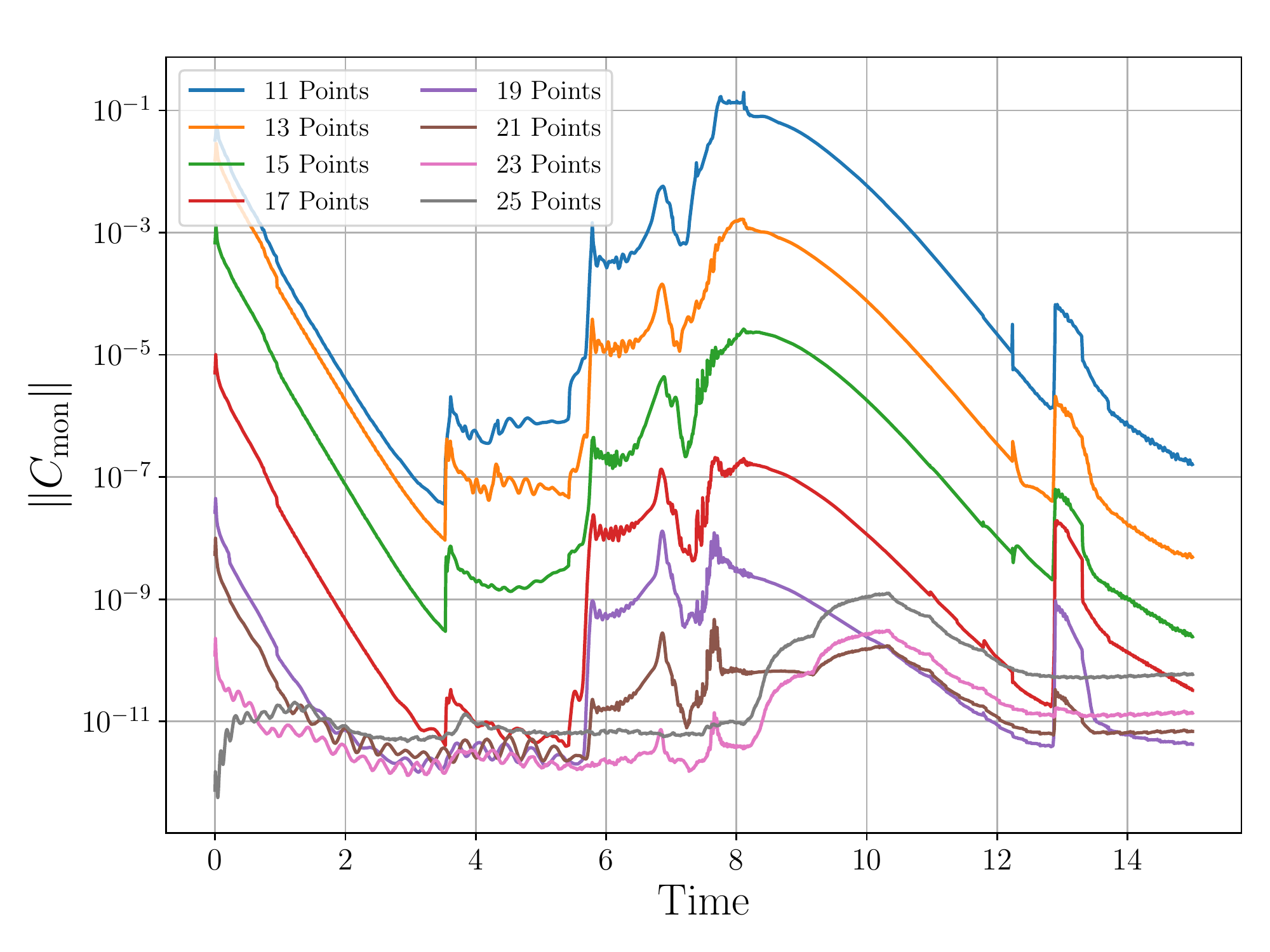}
\caption{Integral of the internal constraint violation monitor during the time evolution of a real scalar field using a fixed h-refinement schedule, with differing static per-grid resolutions.}\label{fig:scalarfield-convergence}
\end{figure}

We find that the constraint violation during early times is larger by around three orders of magnitude when using AMR, compared to the static `average' resolution.
However, once strong features develop in the solution,
constraint violations on the low resolution and average resolution static meshes increase significantly.
The constraint violations of the adaptive grid,
while increasing as well,
stay between five and nine orders of magnitude below those on the static meshes.
Fig.~\ref{fig:scalarfield-error-times-work} shows the total constraint violation over time,
adjusted for cumulative necessary work.
Notably, the highest resolution static mesh shows the lowest constraint violations after the formation of strong features,
staying at a constant level determined by finite floating point precision.
This suggests that in 1d simulations,
using a very high resolution is still feasible,
due to the much lower amount of arithmetic involved in operations on one-dimensional grids,
avoiding the accumulation of large round-off errors that we observe in the two-dimensional examples.
After adjusting the constraint violation for the computational workload,
the high resolution run retains the best efficiency,
as the decreased cost of running with AMR
(in this example the AMR configuration required less than $0.6\%$ of the work needed for the high resolution run)
is not sufficient to make up for the decrease in overall accuracy.

For this example as well we observe exponential convergence as the polymial order of each element is increased, see Fig.~\ref{fig:scalarfield-convergence},
up to the point that saturation is reached and numerical roundoff errors dominate.

\section{Brill Waves}\label{sec:brillwave}

Finally, we consider the case of vacuum gravitational collapse in axisymmetry.
We choose initial data analogous to that used in \cite{SuaRenCor22},
specifically picking an off-center prolate Brill wave
which is known to be subcritical.
We evolve this data using the GHG evolution system (\ref{eq:ghg}), this time with $\alpha\gamma_0=\gamma_2=1$ and using the cartoon method~\cite{AlcBraBru99a} to suppress the angular dimension corresponding to the symmetry.

\begin{figure}[t!]
\includegraphics[width=\columnwidth]{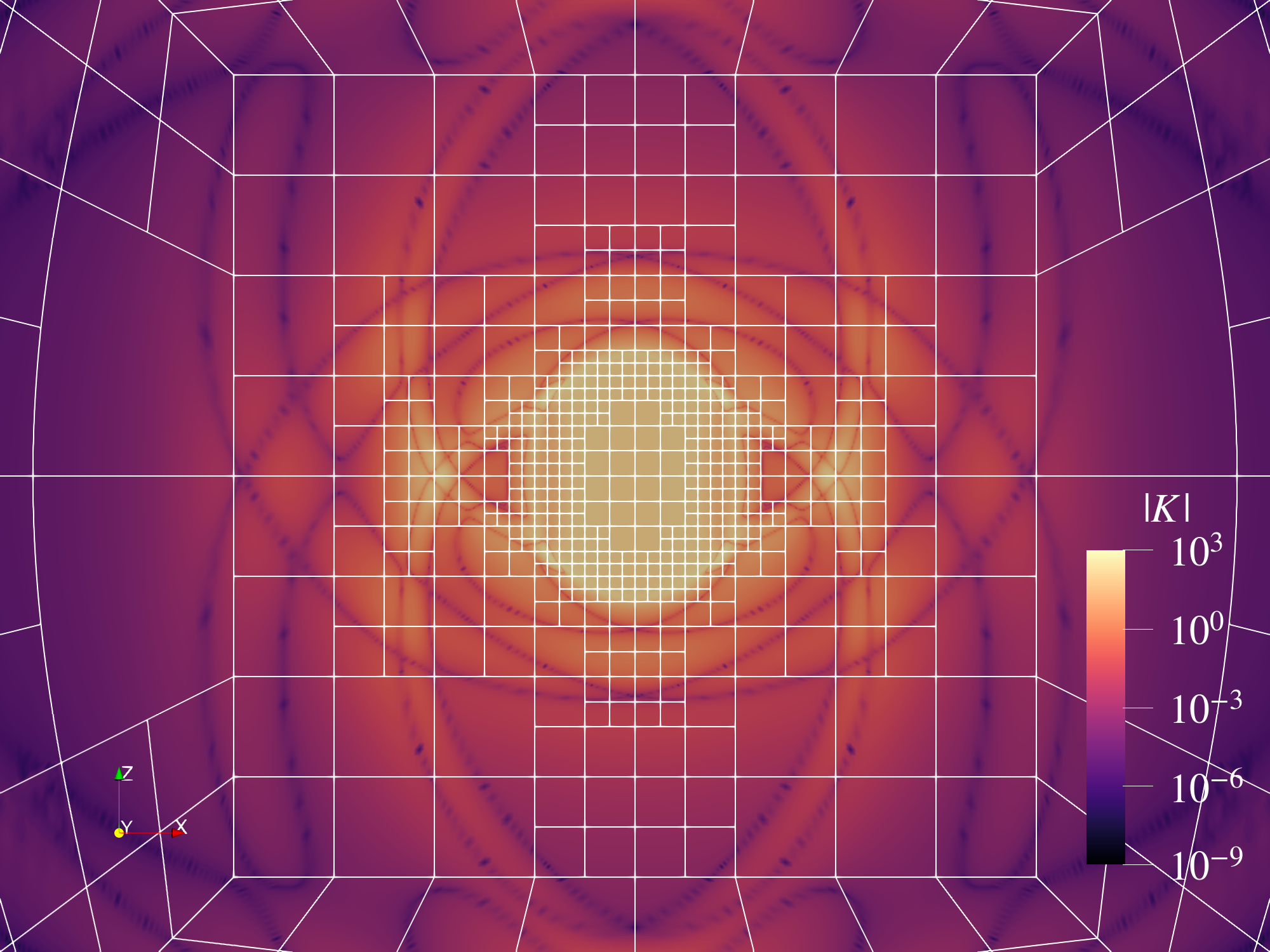}
\caption{Kretschmann scalar $K$ and the grid structure generated by AMR during the time evolution of a Brill wave, based on the `smoothness' heuristic.}
\label{fig:brillwave-refinement}
\end{figure}

\begin{figure}[t!]
\includegraphics[width=\columnwidth]{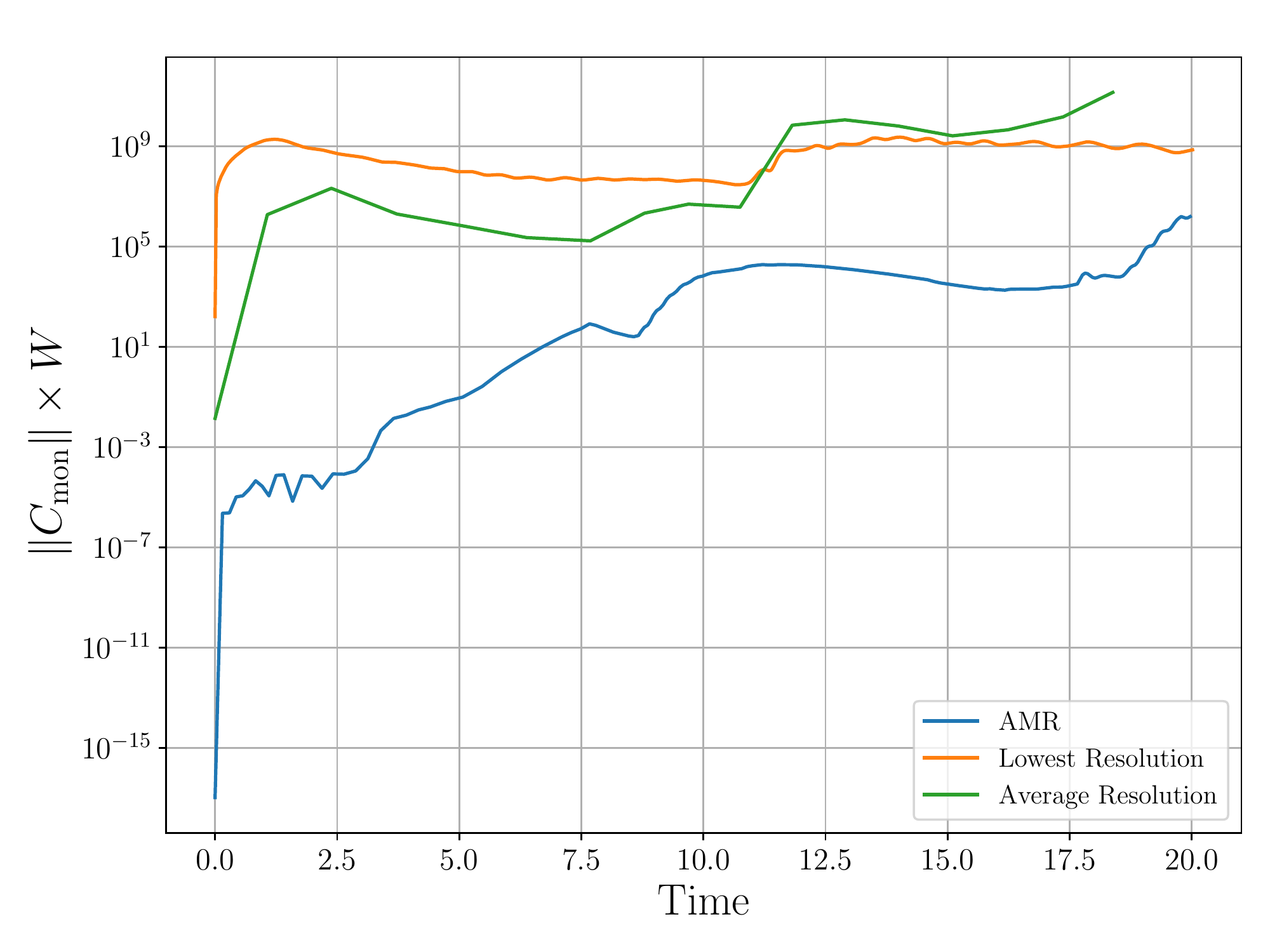}
\caption{Integral of the internal constraint violation monitor, multiplied by the cumulative workload $W$, during the time evolution of a Brill wave using AMR, as compared to the base resolution used by the AMR and a static resolution with comparable total workload.}\label{fig:brillwave-error-times-work}
\end{figure}

To gauge the efficiency of the AMR,
we choose a representative simulation of an off-center Brill wave,
with initial data parameters of $\rho_0 = 5$ and $A = 0.06410$.
This configuration is known to disperse in finite time \cite{SuaRenCor22}.
Fig.~\ref{fig:brillwave-refinement} shows a snapshot of the grid structure as well as the Kretschmann scalar during the time evolution of this data with AMR enabled.
We perform the same comparison as in Sec.~\ref{sec:nonlinear-wave-equation} and Sec.~\ref{sec:scalarfield},
first evolving this initial data with AMR (configuration~\ref{item-amr},
and then determining a configuration of static grids that takes a comparable total workload to evolve (configuration~\ref{item-avg}.
To evolve configuration~\ref{item-lowres},
the per-grid resolution had to be raised from $21 \times 21$ to $23 \times 23$ points,
otherwise the evolution would become unstable after only a short time.
For this comparison,
it was not feasible to also evolve configuration~\ref{item-highres},
as this configuration also quickly developed instabilities.
This may be caused by excessive round-off error accumulation,
similar to the results shown in Fig.~\ref{fig:nonlinear-wave-error-work}.

We again use the constraint monitor variable $C_{\mathrm{mon}}$ as a proxy for numerical error.
In this example, too, the adaptive mesh vastly outperforms the static configuration,
showing more than six orders of magnitude smaller constraint violations.
The total constraint violation adjusted by necessary work is shown in Fig.~\ref{fig:brillwave-error-times-work},
showing the higher efficiency of AMR even as higher overall accuracy is achieved.

\begin{figure}[t!]
\includegraphics[width=\columnwidth]{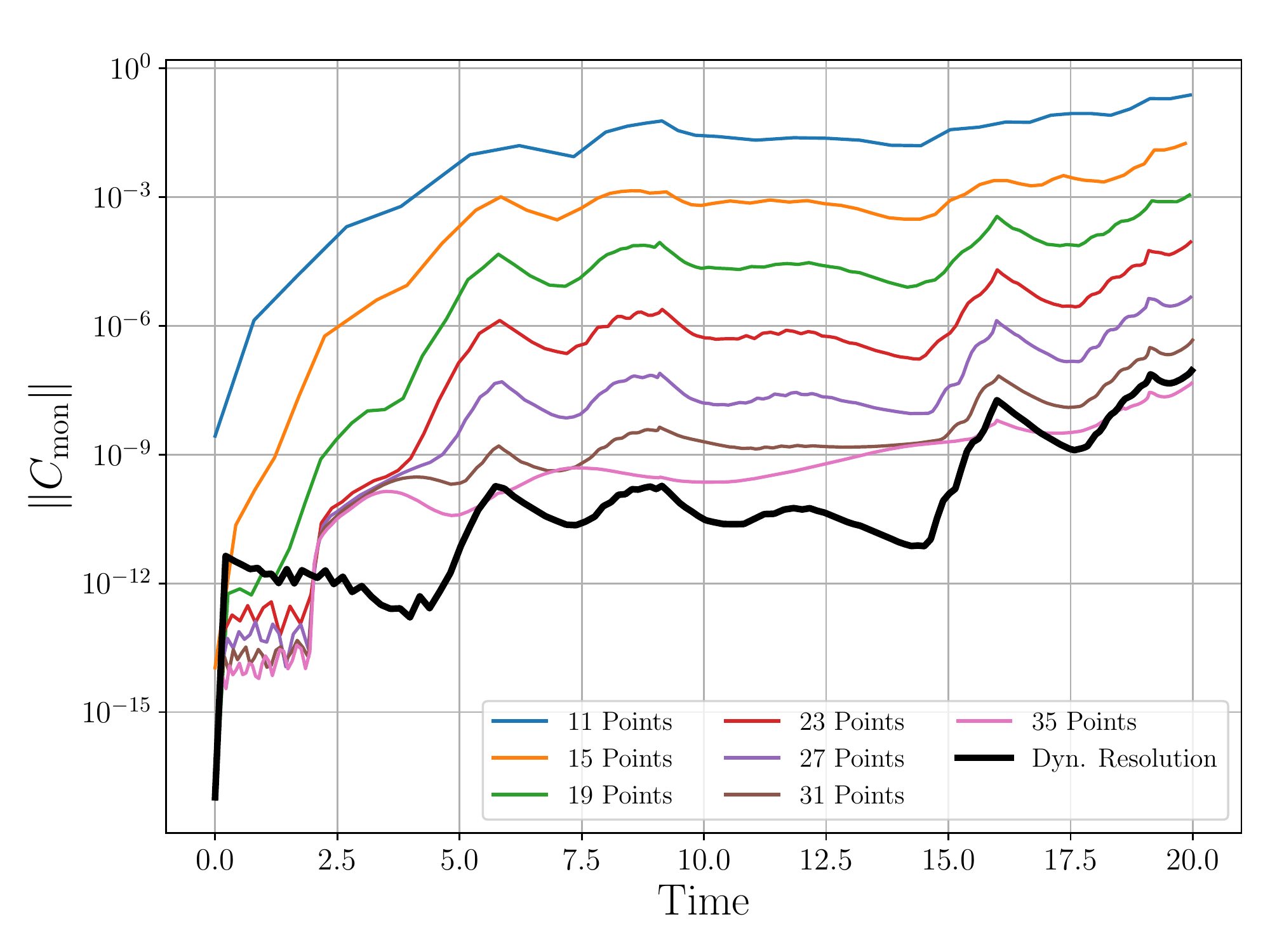}
\caption{Integral of the internal constraint violation monitor during the time evolution of a Brill wave using a fixed h-refinement schedule, with differing static per-grid resolutions, as well as a dynamic resolution determined by the p-refinement scheme.}\label{fig:brillwave-convergence}
\end{figure}

To verify the convergence of the method,
we evolve identical initial data on several different per-grid resolutions.
To make the results comparable,
the h-refinement is in each case driven by a predetermined refinement schedule,
generated by a reference run using $21 \times 21$ points per grid.
We again use the integral of the constraint monitor $C_{\mathrm{mon}}$ as a proxy for numerical error.
Fig.~\ref{fig:brillwave-convergence} shows the integrated constraint violations for a variety of resolutions, showing exponential convergence of the solution.

Also in Fig.~\ref{fig:brillwave-convergence}, we show the constraint violations for a run with dynamic, local p-refinement.
For this, we use the same fixed h-refinement schedule as for the static resolutions,
and additionally enable p-refinement,
using the truncation error estimate,
evaluated on all metric components, as well as $C_{\mathrm{mon}}$ itself.
This results in overall constraint violations that are at times lower than the highest static resolution of $35 \times 35$ points per grid,
despite this also being the highest resolution available to the AMR system.
By the end of the evolution the constraint violations are comparable between high resolution static and dynamic resolutions.
However, evolving the system with p-refinement enabled requires only roughly $22\%$ of the computational work as evolving with static $35 \times 35$ points per grid,
demonstrating the much higher efficiency obtained by dynamic p-refinement.

\section{Conclusions}

We have succesfully implemented fully adaptive hp-refinement into the \texttt{bamps} code,
and demonstrated the spectral convergence of simulations on the grid structures generated by adaptive h-refinement,
as well as the gains in accuracy and efficiency obtained through the use of p-refinement.
Using this new system,
numerical evolutions for several physical systems show improvements in accuracy by several orders of magnitude,
when compared to evolutions using static resolutions requiring similar amounts of work.

In the case of two-dimensional evolutions,
the adaptive resolutions show increased numerical efficiency,
that is less work is needed to reach a specific accuracy,
even while greater total accurary is achieved.
This does not hold true, however,
for the one-dimensional example we studied.
Here, while using an adaptive resolution is computationally cheaper by several orders of magnitude,
these savings do not make up for the decrease in overall accuracy.

It must be noted that the proxy used to measure numerical error,
the integral of overall constraint violations,
is an imperfect indicator of overall accuracy.
We also observe that using very high resolutions is not feasible in some cases,
as the accumulated round-off errors due to large amounts of arithmetic performed in the course of matrix multiplications destroys the accuracy of the simulation.

We find that \texttt{bamps} shows near-perfect scaling up to 1000 CPU cores,
and satisfactory scaling up to 6000 cores.
Current production runs using \texttt{bamps} in 1d and 2d do not exceed 1000 CPU cores.
While we focused here on 1d and 2d tests, the design of
\texttt{bamps} is aimed at fully 3-dimensional simulations.
In 3d, the ratio of overhead to work is more favorable, in particular for scaling,
which we have confirmed in preliminary tests.

A previous version of \texttt{bamps} only supporting h-refinement has
already been used successfully to further the study of critical
collapse of gravitational waves \cite{SuaRenCor22}, and the full
hp-refinement algorithm is currently being applied in simulations of
both the critical collapse of real scalar fields, as well as the time
evolution of complex scalar fields. In particular, the evolution of
boson stars represents an ideal use case of the methods presented
here. These sample applications focus on smooth fields.
AMR is also a de-facto necessary feature to effectively study problems
involving general relativistic hydrodynamics (GRHD).
However, more work is necessary to fully manage emerging shocks and
other discontinuities in conjunction with AMR (see \cite{BugDieBer15}
for an exploration of GRHD in \texttt{bamps} using non-adaptive
meshes).
A technique that remains to be implemented is that of local time
stepping to obtain hpt-refinement, where elements of different
resolutions are advanced in time at different rates, which offers
great potential for increased efficiency. Spectral element methods are
well-suited to local time stepping due to the discontinuous coupling
of elements. Local time stepping schemes generally require significant
changes to the underlying time evolution infrastructure, which are
currently underway.

The technical upgrade of \texttt{bamps} will benefit most, if not all,
future projects using the code. Furthermore, the methods developed for
this purpose and the insights gained here can serve as a case study
with a wide range of applicability.

\acknowledgments 
We are grateful to F. Atteneder, H. R. Rüter, and I. Su\'arez Fern\'andez
for helpful discussions and for collaboration on other aspects of
\texttt{bamps}.
This work was partially supported by the FCT (Portugal) IF
Program~IF/00577/2015, Project~No.~UIDB/00099/2020 and
PTDC/MAT-APL/30043/2017,
and in part by the Deutsche Forschungsgemeinschaft (DFG) under Grant
No.\ 406116891 within RTG 2522/1 and DFG Grant BR 2176/7-1.

\normalem
\bibliography{hp_amr.bbl}

\end{document}